\documentclass[aip,reprint,groupedaddress,amsmath,amssymb,longbibliography]{revtex4-2}

\usepackage{CJK}
\usepackage{ulem}
\usepackage{graphicx} % Include figure files
\usepackage{bm}% bold math
\usepackage[pdfstartview=FitH, CJKbookmarks=true, bookmarksnumbered=true, bookmarksopen=true, colorlinks=true, pdfborder=0 0 1, citecolor=blue, linkcolor=blue, linktocpage=true] {hyperref}
\usepackage{epstopdf} % ensure that the PDFlatex can work normally.

\begin{document}

\title{Probing Chiral-Symmetric Higher-Order Topological Insulators with Multipole Winding Number}
% Real-space characterization of winding number from the bulk in chiral-symmetric higher-order topological insulators
% (order of authors and affiliations are tentative) 

\author{Ling Lin$^{1,2}$}
\author{Chaohong Lee$^{1,2}$}
\email{Corresponding author. Email: chleecn@szu.edu.cn, chleecn@gmail.com}

\affiliation{$^{1}$Institute of Quantum Precision Measurement, State Key Laboratory of Radio Frequency Heterogeneous Integration, College of Physics and Optoelectronic Engineering, Shenzhen University, Shenzhen 518060, China}
\affiliation{$^{2}$Quantum Science Center of Guangdong-Hong Kong-Macao Greater Bay Area (Guangdong), Shenzhen 518045, China}

\date{\today}

%=========Abstract=========

\begin{abstract}
The interplay between crystalline symmetry and band topology gives rise to unprecedented lower-dimensional boundary states in higher-order topological insulators (HOTIs).
However, the measurement of the topological invariants of HOTIs remains a significant challenge.
Here, we define a {multipole winding number} (MWN) for chiral-symmetric HOTIs by applying a corner twisted boundary condition. 
The MWN, arising from both bulk and boundary states, accurately captures the bulk-corner correspondence  including boundary-obstructed topological phases.
To address the measurement challenge, we leverage the perturbative nature of the corner twisted boundary condition and develop a real-space approach for determining the MWN in both two-dimensional and three-dimensional systems.
The real-space formula provides an experimentally viable strategy for directly probing the topology of chiral-symmetric HOTIs through dynamical evolution.
Our findings not only highlight the twisted boundary condition as a powerful tool for investigating HOTIs, but also establish a paradigm for exploring real-space formulas for the topological invariants of HOTIs.
\end{abstract}

% ========Main body=========================================================
\maketitle
% \tableofcontents  % optional

\section*{Introduction}
In contrast to conventional topological insulators, higher-order topological insulators (HOTIs)~\citep{sciadv.aat0346,PhysRevLett.119.246402,PhysRevLett.119.246401} are characterized by higher-order topological invariants, resulting in the presence of lower-dimensional boundary states. 
The emergence of these boundary states is typically associated with non-trivial higher-order bulk topology~\citep{xie2021higher}.
Recent experiments have successfully demonstrated the realization of HOTIs~\citep{serra2018observation,peterson2018quantized,xue2020observation,ni2020demonstration,kempkes2019robust,mittal2019photonic,PhysRevLett.126.146802,schulz2022photonic}, as evidenced by the appearance of corner states. 
However, the measurement of topological invariants of HOTIs still remains a significant challenge, as topological invariants are a global quantity, which are usually related to the bulk.

In addition to exploring the edge topology, it is crucial to characterize bulk topology.
In one-dimensional (1D) chiral-symmetric topological insulators, it is known that the winding number can be obtained by measuring mean chiral displacement in the bulk~\citep{Science362, PhysRevA.98.013835, xie2019topological, PhysRevLett.122.193903,PhysRevLett.123.080501,PhysRevResearch.2.023119,PhysRevLett.124.050502}.
This approach is based on the real-space formula of the winding number~\citep{PhysRevLett.113.046802,Maffei2018}.
Recently, the multipole chiral number (MCN) approach is proposed to capture the topology of chiral-symmetric HOTIs based upon the difference in the multipole moments of two sublattices in real space~\citep{PhysRevLett.128.127601}.
Chiral symmetry has been demonstrated to safeguard non-trivial higher-order topology independently of crystalline symmetry~\citep{PhysRevLett.125.166801, PhysRevB.103.085408,PhysRevLett.128.127601,PhysRevLett.131.157201,PhysRevB.109.014204}.
The MCN represents a generalization of the real-space winding number for conventional topological insulators \citep{PhysRevB.103.224208} and effectively characterizes the higher-order topology. 
Remarkably, the MCN is a $\mathbb{Z}$ number for characterizing chiral-symmetric topological insulators.
This is in stack contrast to the multipole polarization, which only gives rise to a $\mathbb{Z}_2$ quantity. 

Up to now, the experimental investigation of the chiral-symmetric HOTIs with larger topological numbers relies on identifying the corner states \citep{PhysRevLett.131.157201,PhysRevB.108.205135,PhysRevApplied.20.064042}.
To probe the bulk topology of HOTIs, it has been proposed to generalize the concept of mean chiral displacement.
By combining the winding numbers in $x$ and $y$ directions~\citep{OE:Lu:20}, the method of measuring mean chiral displacement for 1D systems can be extended to 2D chiral-symmetric HOTIs.
However, this method remains within the realm of conventional topological insulators. 
Besides, a generalization of the mean chiral displacement to the quadrupole insulator~\citep{PhysRevLett.126.016802}  successfully extracts the quadrupole nature of the chiral-symmetric HOTI from the bulk.
Given the similarities between the first-order chiral-symmetric topological insulator and the higher-order HOTI, a question arises: Can we define a winding number for chiral-symmetric HOTIs?
Furthermore, can we directly measure such a winding number?

In this article, we present a comprehensive and experimentally viable framework for probing chiral-symmetric HOTIs using a real-space topological invariant.
We define the {multipole winding number} (MWN), a concept lacking a momentum-space counterpart, based on the bulk-corner correspondence.
This topological invariant is defined through the corner twisted boundary condition (CTBC), linking corners and addressing gauge fields during tunneling across different corners.
It precisely quantifies the number of zero-energy modes in both 2D and 3D chiral-symmetric HOTIs under open boundary conditions even for the boundary-obstructed topological phase (BOTP).
Remarkably, the MWN and the corner modes are found to be contributed from both bulk and boundary states.
Similar to deriving real-space winding numbers for 1D chiral-symmetric topological insulators~\citep{PhysRevB.103.224208} and real-space Chern numbers for 2D topological insulators~\citep{PhysRevB.108.174204}, we develop a real-space formula for the MWN via perturbative expansion.
We elaborate appropriate twist operators that transform the gauge field at corners into the bulk, rendering the gauge field a reliable perturbative parameter.
In further, we demonstrate numerically how to extract the MWN through dynamical evolution. 
Our proposed scheme can be easily implemented using currently available experimental techniques.

\section*{Results}

% \subsection*{Corner twisted boundary condition and multipole winding number}
%
\textbf{Corner twisted boundary condition and multipole winding number.}
It is known that chiral-symmetric topological insulators in odd dimensions are characterized by the winding number~\citep{RevModPhys.88.035005}.
In 1D, the winding number can be regarded as the dipole polarization between the two sublattices~\citep{PhysRevB.103.224208}.
With a non-trivial winding number in bulk, zero-energy edge states emerge under OBC, which is called the bulk-edge correspondence.
To capture the dipole quantity of the bulk, one can introduce a general twisted boundary condition (TBC)~\citep{PhysRevB.74.045125, PhysRevB.103.224208}.
% 
% To prove this bulk-edge correspondence, one can introduce a parameter $r$ on the tunneling across the boundary~\citep{PhysRevB.74.045125}.
%
% This parameter continuously connects the OBC and the TBC, and thus the winding number is related to the closed line integral of $z\equiv \det h(r,\theta) \in \mathbb{C}$,
By introducing a parameter $r$ on the tunneling across the boundary~\citep{PhysRevB.74.045125}, which connects the OBC and the TBC, the winding number $\nu$ is determined by the closed line integral of $z\equiv \det h(r,\theta) \in \mathbb{C}$,
\begin{eqnarray}
\label{eqn:winding_number_z_complex}
\nu  = \frac{1}{{2\pi i}} \oint\nolimits_{\left( {r,\theta } \right) \in \mathcal{C}} {{\rm{d}}\theta \;{\rm{Tr}}\left[ {h{{\left( {r,\theta } \right)}^{ - 1}}{\partial _\theta }h\left( {r,\theta } \right)} \right]} ,
\end{eqnarray}
which requires $h$ to be invertible (i.e. all its singular values are non-zero) on the path.
The winding number $\nu$ reflects the fold of zero points $(z=0)$ encircled by the curve $\mathcal{C}$ on the complex plane.
Meanwhile, since $\det h(r,\theta)  \propto \prod\nolimits_n {{E_n}} $ (see Methods for proof), these zero points are related to the number of zero singular values, which is half of the number of zero-energy modes $N_{E=0}$, i.e. $N_{E=0} = 2\nu$.
Thus, the winding number defined through TBC can well characterize the number of zero-energy edge states.
%
% Generally, this only occurs if $r=0$, that is, the system is under OBC or the system undergoes a phase transition.
%
% Of course, the zero point can also appear elsewhere ($r\ne0$), and we shall circumvent this case by choosing an appropriate TBC.

In a HOTI, the total dipole moment vanishes, while the higher-order moment can be non-zero.
Usually, a HOTI is protected by crystalline symmetry.
In the presence of chiral symmetry, a HOTI can be classified by integers and its zero-energy modes are robust against weak disorders~\citep{PhysRevLett.128.127601}, and therefore the crystalline symmetry is not necessary.
In particular, zero-energy modes appear pairwise in a non-trivial chiral-symmetric HOTI.
However, there is no well-defined momentum-space formula for determining the topological invariant protected by chiral symmetry.
To characterize the topology of chiral-symmetric HOTIs, we use a bottom-to-up strategy starting from characterizing non-trivial zero-energy modes.
The essence of this approach lies in imposing a {generalized TBC} to eliminate zero-energy modes, and then use a parameter $r \in[0, 1]$ to continuously connect this generalized TBC to OBC.
This allows us to define the winding number on the complex plane $z\equiv \det h(r, \theta)$ to extract the number of singular points (zero-energy modes).
The generalized TBC should keep $\det h(r, \theta) \ne 0$ for all $r\ne 0 $ and $\theta \in [0, 2\pi]$, ensuring that zero-energy modes only appear under the OBC ($r=0$).
Within these configurations, one can still use Eq.~\eqref{eqn:winding_number_z_complex} to calculate the topological invariant of HOTIs.

\begin{figure}
\includegraphics[width = \columnwidth ]{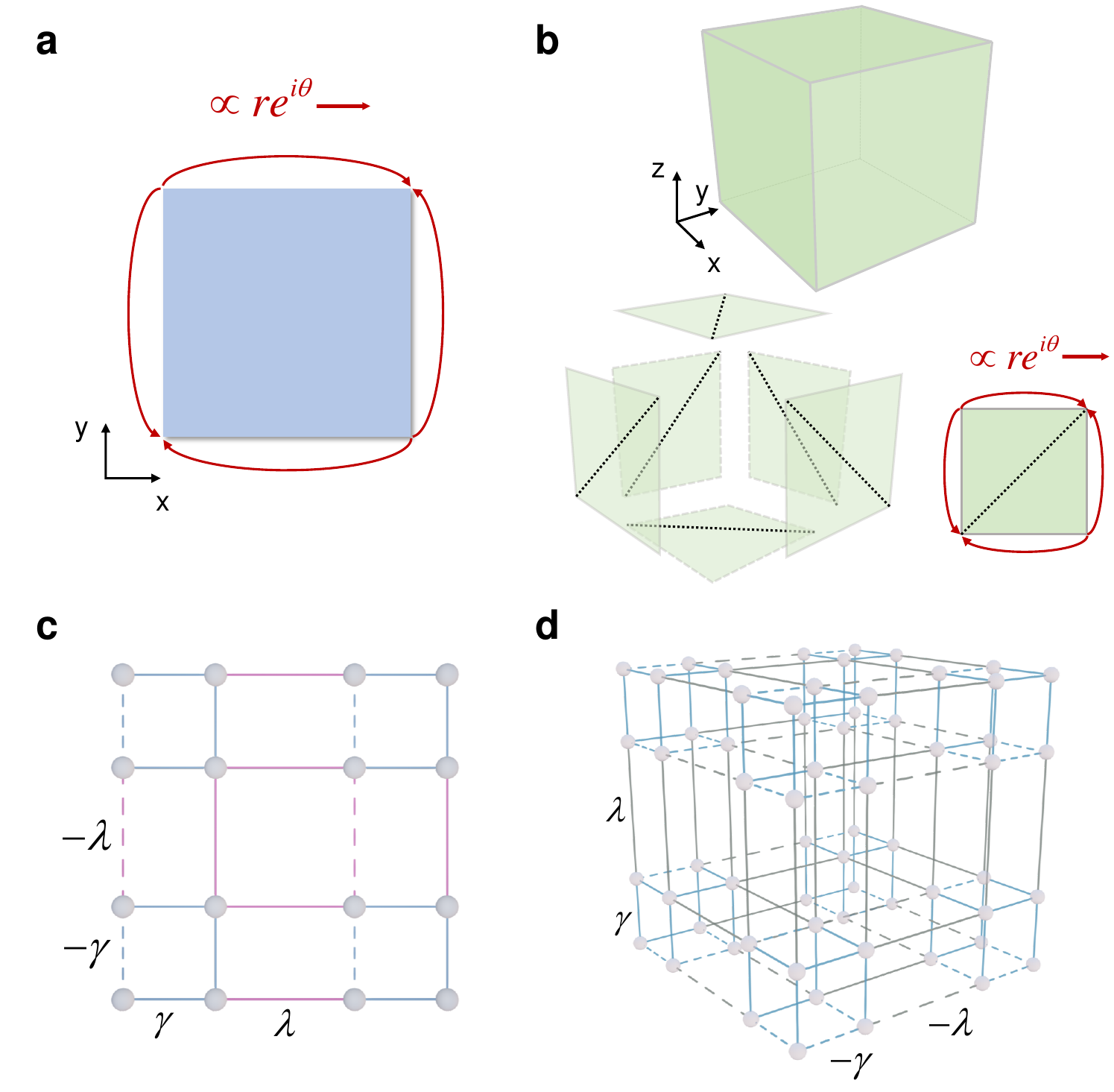}
\caption{\label{fig:FIG_BBH_Model}
\textbf{Schematic for corner twisted boundary condition and the Benalcazar-Bernevig-Hughes model.}
\textbf{a} 2D case. The red arrows indicate the tunneling from one corner to the other, which are parametrized by $r \in [0, 1]$ and $\theta \in [0, 2\pi]$. 
\textbf{b} 3D case. The CTBC is imposed on each face, with the dashed line indicating the orientation of the twist angle.
\textbf{c-d} 2D and 3D Benalcazar-Bernevig-Hughes model under OBC.
The tunneling amplitudes $\gamma$ and $\lambda$ are marked by different colors.
The dashed lines represent negative amplitudes.
}
\end{figure}

To meet the aforementioned criteria, we consider a {corner twisted boundary condition} (CTBC) shown in Fig.~\ref{fig:FIG_BBH_Model} (a-b), in which particles are allowed to tunnel between corners with a complex amplitude.
In 2D square lattices, we adopt the CTBC~\citep{PhysRevLett.128.246602}.
In 3D cubic lattices, we consider the corner tunneling in each face of the cube, in which the orientation of the CTBC is implied by the dashed line.
The CTBC is equivalent to threading a gauge field, resembling the general TBC.
This choice of gauge field ensure that it will not significantly change the spectrum, which is confirmed by perturbative analysis later.
To manifest the bulk-corner correspondence, the tunneling strength between corners is required to multiply a factor $r\in[0,1]$.
%, and only the parameter $r$ will induce zero-energy modes from the gapped bulk.
%
With these conditions, we can define the MWN via Eq.~\eqref{eqn:winding_number_z_complex}.
In addition, the CTBC can be used to define the multipole polarization~\citep{PhysRevLett.128.246602}, which is quantized by the chiral symmetry~\citep{PhysRevLett.125.166801, PhysRevB.103.085408} (see proof in Sec. A of Supplementary Note 1.).
%

\begin{figure}
  \includegraphics[width = \columnwidth ]{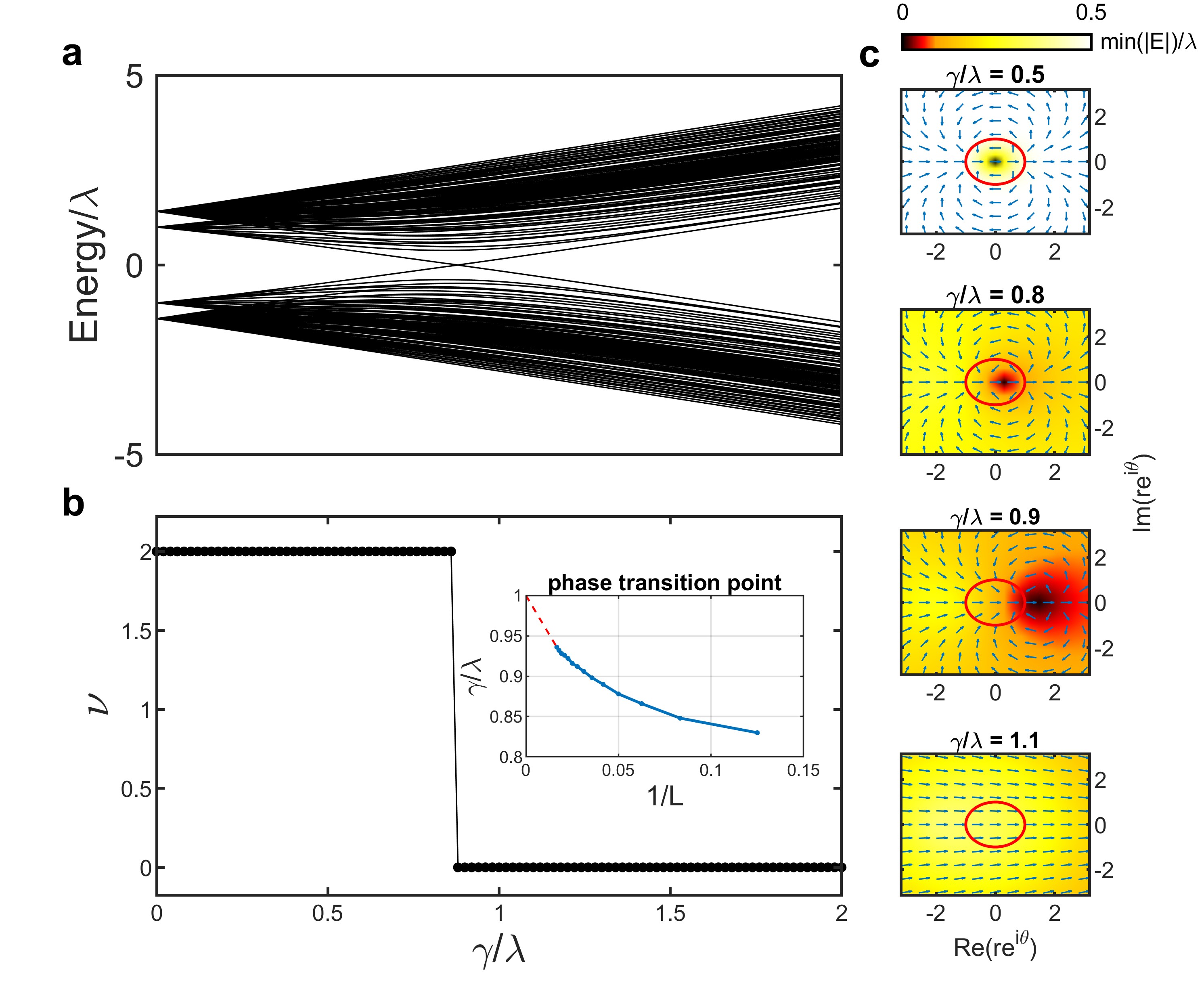}
\caption{\label{fig:FIG_Winding_Number_BBH_2D}
  \textbf{Energy spectrum and the MWN of the 2D BBH model and schematic illustration of the winding number.}
  \textbf{a} Energy Spectrum computed from exact diagonalization as a function of $\gamma/\lambda$ under the CTBC.
  \textbf{b} The corresponding MWN.
  Inset in \textbf{b}: the critical $\gamma/\lambda$ versus $1/L$ with a red dashed line for guidance.
  \textbf{c} The lowest singular values $\min(\Sigma)$ of $h(r, \theta)$ on the complex plane for $L_x=L_y=L=20$.
  Arrows represent the angle of $\det h(r, \theta)$.
    }
\end{figure}

% \subsection*{Benalcazar-Bernevig-Hughes model}
%%
\textbf{Benalcazar-Bernevig-Hughes model.}
To illustrate the validity of our MWN, we consider the Benalcazar-Bernevig-Hughes (BBH) model~\citep{science.aah6442} (see Fig.~\ref{fig:FIG_BBH_Model} (c-d)).
This model preserves chiral symmetry and exhibits zero-energy corner states from higher-order topology.
Notably, the BBH model with long-range tunneling has been realized in experiments~\citep{PhysRevLett.131.157201,PhysRevB.108.205135,PhysRevApplied.20.064042}.
In 2D, the system obeys the tight-binding Hamiltonian: 
\begin{equation}
\label{eqn:2D_BBH_model}
    {{\hat H}_{{\rm{2D}}}} =  - \sum\limits_{x,y} {\left[ {{t_x}\hat a_{x + 1,y}^\dag {{\hat a}_{x,y}} + {{\left( { - 1} \right)}^x}{t_y}\hat a_{x, y + 1}^\dag {{\hat a}_{x,y}}} \right]}  + {\rm{H.c.}},
\end{equation}
with ${t_x} = \frac{{\gamma  + \lambda }}{2} + \frac{{\gamma  - \lambda }}{2}{\left( { - 1} \right)^x},{t_y} = \frac{{\gamma  + \lambda }}{2} + \frac{{\gamma  - \lambda }}{2}{\left( { - 1} \right)^y}$.
It is known that there are two topological phases determined by the ratio $|\gamma/\lambda|$.
If $|\gamma/\lambda|<1$, a non-zero bulk multipole polarization arises and zero-energy modes emerge at corners under the OBC.
Otherwise, if $|\gamma/\lambda|>1$, the system is well gapped and there is no in-gap zero modes.
From the multipole polarization under periodic boundary condition (PBC), the phase transition point can be given as $|\gamma/\lambda|=1$.
Under the CTBC, we calculate the spectrum and the MWN, see Fig.~\ref{fig:FIG_Winding_Number_BBH_2D} (a-b).
Interestingly, the phase transition point, which is characterized by the gap closing and the sudden change of MWN, deviates from $\gamma/\lambda = 1$.
This can be attributed to the finite-size effect.
Under OBC, the localization length of corner states near the phase transition point becomes significantly large, leading to hybridization of corner states and resulting in a non-zero energy shift in finite systems.
According to Eq.~\eqref{eqn:winding_number_z_complex}, the MWN is related to the number of zero-energy mode under OBC inside the integral circle ($r=1$).
Near the phase transition point, the singular point gradually move out of the integral circle due to the energy shift, causing a change of MWN.
Thus, the phase transition point given by the MWN is also shifted due to the finite-size effect (see detailed discussions in Sec. B of Supplementary Note 2).
In the thermodynamic limit $L\to \infty$, the phase transition point approaches to $\gamma/\lambda = 1$, as shown in the inset of Fig.~\ref{fig:FIG_Winding_Number_BBH_2D} (b).
This shows that the MWN can well capture the bulk topology.

Subsequently, we calculate the lowest non-negative energy $\min({|E|})$ and the determinant $\det h(r,\theta)$ as functions of $r$ and $\theta$, see the right panel of Fig.~\ref{fig:FIG_Winding_Number_BBH_2D}.
%
% The singular values of $h(r,\theta)$ correspond to the positive eigenenergies of $H$, and the eigenenergies appear pair-wisely with opposite values.
%
% The lowest singular value of $h(r,\theta)$ corresponds to the eigenenergy most close to $E=0$.
%
For $|\gamma/\lambda|<1$ (and away from the phase transition point), we confirm that there appears $2$ ($4$)-fold degenerate singular points at $r=0$ in the 2D (3D) BBH model.
These singular points can be viewed as topological defect, which are precisely captured by the MWN and consistent with half the number of corner states under OBC.
%
% This is caused by the finite-size effect.

\begin{figure}
  \includegraphics[width = \columnwidth ]{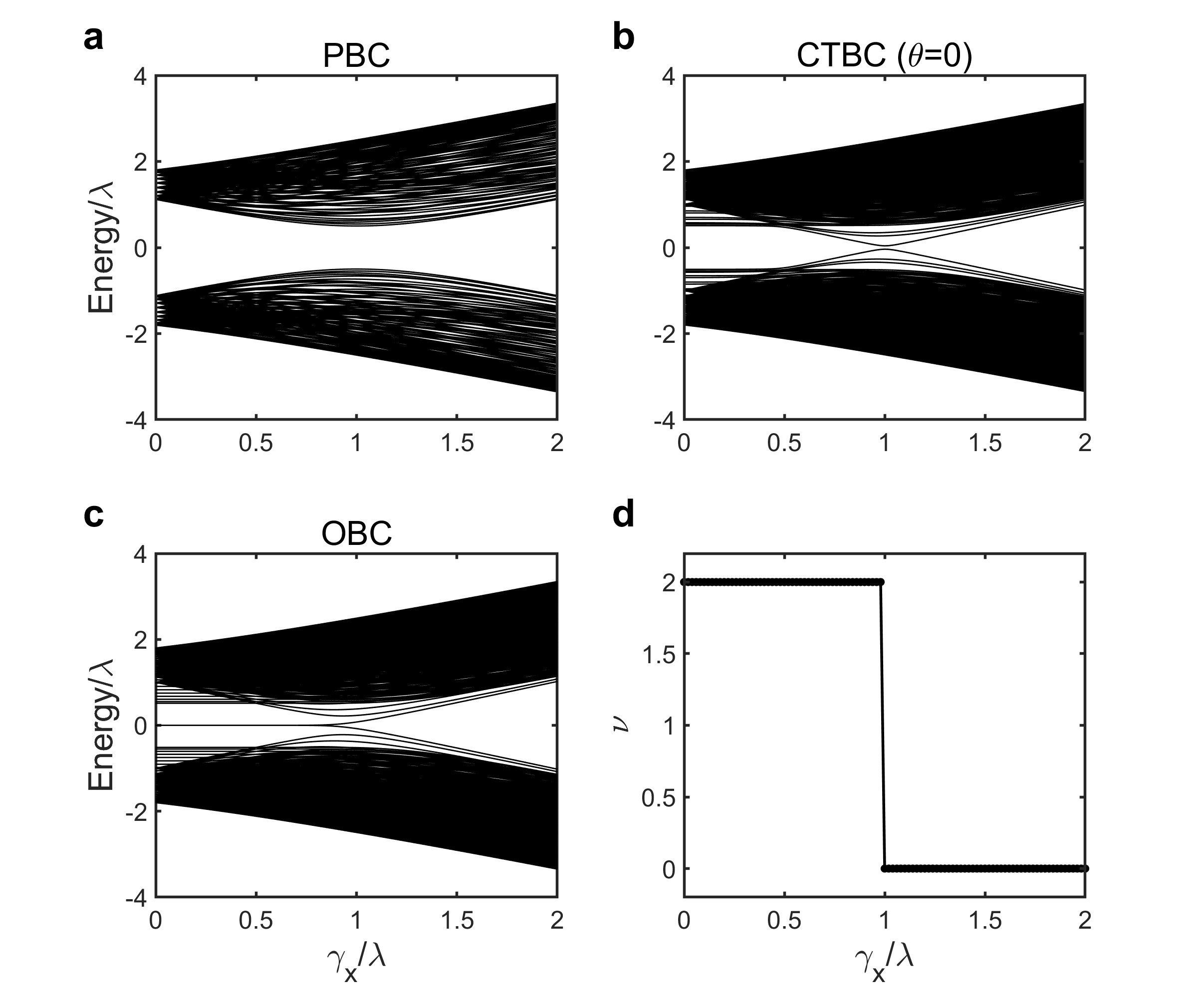}
\caption{\label{fig:FIG_BOTP}
  \textbf{Energy spectrum and the MWN of the 2D BBH model under the boundary obstructed topological phase (BOTP).}
  \textbf{a-c} Energy Spectrum computed from exact diagonalization under the PBC, CTBC, and OBC respectively.
  \textbf{d} The corresponding MWN computed via Eq.~\eqref{eqn:winding_number_z_complex} under CTBC.
  Other parameters are set to: $\gamma_y/\lambda=0.5$ and $L=40$.
  }
\end{figure}

%\subsection*{Boundary-obstructed topological phase}
%
\textbf{Boundary-obstructed topological phase.}
Furthermore, we demonstrate that the MWN can effectively characterize the topology and the bulk-corner correspondence for boundary-obstructed topological phase (BOTP). 
It has been shown that different BOTPs can be deformed continuously to each other without closing the bulk gap under the PBC \citep{PhysRevResearch.3.013239}.
Nevertheless, the phase transition between different BOTPs is associated with gap closing under the OBC.
To realize a BOTP, we break the $C_4$ symmetry in the 2D BBH model by letting ${t_x} = \frac{{\gamma_x  + \lambda }}{2} + \frac{{\gamma_x  - \lambda }}{2}{\left( { - 1} \right)^x},{t_y} = \frac{{\gamma_y  + \lambda }}{2} + \frac{{\gamma_y  - \lambda }}{2}{\left( { - 1} \right)^y}$ in Eq.~\eqref{eqn:2D_BBH_model}.
In this model, zero-energy modes emerge under OBC when $|\gamma_{x,y}/\lambda|<1$.
If either $|\gamma_{x}/\lambda|>1$ or $|\gamma_{y}/\lambda|>1$, no zero-energy mode exists, indicating that $|\gamma_{x,y}/\lambda|=1$ are phase boundaries. 
However, under PBC, the spectral gap only closes at some points where $|\gamma_{x}/\lambda|=|\gamma_{y}/\lambda|=1$.
Thus, the phase transition between BOTPs can not be characterized by the gap closing of the bulk.
Instead, by employing CTBC, we observe a distinct gap-closing point nearer either $|\gamma_{x}/\lambda|=1$ or $|\gamma_{y}/\lambda|=1$, as shown in Fig.~\ref{fig:FIG_BOTP}.
Remarkably, the MWN accurately captures the presence of zero-energy modes and the phase transition of BOTP under OBC.
This is because the gap closing is resulted not only from the bulk states but also the boundary states under CTBC.
%
% there are both the bulk states and boundary states when $|\gamma_{x,y}|<1$ under CTBC.
%
The non-trivial MWN is contributed from both of them when $|\gamma_{x,y}|<1$ .
Therefore, it can be inferred that the corner state is resulted  from the non-trivial topology of both the bulk and boundary, which will be examined in real space below.

%\subsection*{Real-space representation of multipole winding number}
%
\textbf{Real-space representation of multipole winding number.}
In experiments, the simplest method to distinguish chiral-symmetric HOTIs from trivial topological insulators is through observing their corner states.
In addition to exploring corner states resulted from non-trivial bulk topology, researchers are also making efforts to extract the topological invariant of HOTIs experimentally \citep{PhysRevB.99.045441,OE:Lu:20, PhysRevB.102.094305, PhysRevLett.126.016802}.
Below, we derive a real-space formulation for MWN and explore feasible method for corresponding experimental measurement.

\begin{figure}
\centering
  \includegraphics[width = \columnwidth ]{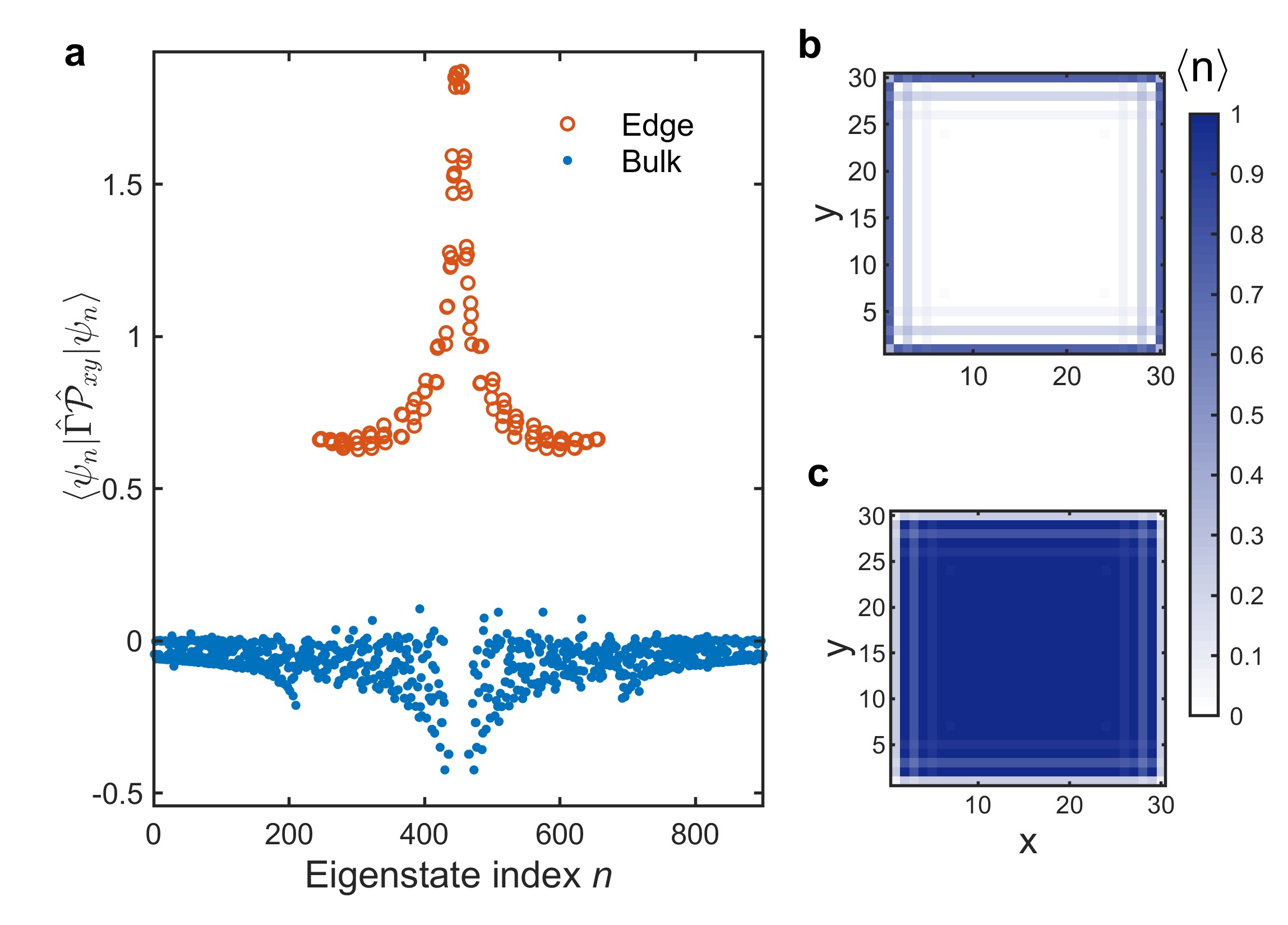} 
\caption{\label{fig:FIG_P_bulk_edge}
  \textbf{Contribution of each eigenstates to the MWN in 2D BBH model.}
  \textbf{a} The expectation of the chiral multipole operator $\langle\psi_n|\hat{\Gamma} \hat{\mathcal{P}}_{xy}|\psi_n\rangle $ for each eigenstate except for zero-energy modes in 2D BBH model.
  Red circles and blued dots are distinguished by its value.
  The indices of eigenstates are arranged in ascending order of energy.
  \textbf{b-c} respectively shows the total density distribution of eigenstates marked by red and blue colors in (\textbf{a}).
  Parameters are chosen as $\gamma/\lambda = 0.5$ and $L=30$.
   }
\end{figure}

To attain the real-space formula, our central idea is to make use of the gauge freedom of the CTBC and find appropriate gauges that makes the twist angle distributed uniformly in the bulk \citep{PhysRevB.103.224208,PhysRevB.108.174204}.
Then the twist angle can be treated as a perturbation in order of $\theta/L$ and one can use the perturbative expansion to derive the MWN.
To perform the desired gauge transformation, we introduce the twist operators
\begin{equation}
\label{eqn:twist_operator_2D}
\hat U_\theta ^{\mathrm{2D}} = \exp \left( {i\frac{\theta }{L}{{\hat {\cal P}}_{xy}}} \right)~\textrm{with}~ {\hat{\cal P}_{xy}} = \sum\limits_{x,y} {|x - y|{{\hat n}_{x,y}}}
\end{equation}
for 2D systems and
\begin{eqnarray}
\label{eqn:twist_operator_3D}
\hat U_\theta ^{\mathrm{3D}} &=& \exp \left( {i\frac{\theta }{L}{{\hat {\cal P}}_{xyz}}} \right) ~\textrm{with}~ {{\hat {\cal P}}_{xyz}} = \sum\limits_{x,y} {{f_{xyz}}{{\hat n}_{x,y}}} , \nonumber \\
{f_{xyz}} &=& \left\{ {\begin{array}{*{20}{c}}
{|x - y| - z + 1,}&{|x - y| \ge z - 1}\\
{|L - x - y| + z - L,}&{|L - x - y| \ge L - z}\\
{0\quad }&{{\rm{otherwise}}}
\end{array}} \right.  \nonumber \\
\end{eqnarray}
for 3D systems.
Through the transformation $\tilde H\left( \theta  \right) = \hat U_\theta \hat H\left( \theta  \right){ {\hat U_\theta } ^{ - 1}}$, the gauge fields at corners are transformed into the bulk with diluted and uniform strength (see illustrations in Sec. C of Supplementary Note 1).
%
%In a 3D system, the gauge field is only distributed near the corners.
%
The Hamiltonian $\hat H\left( \theta  \right)$ under the CTBC satisfies the gauge-invariant relation: $\hat H\left( {\theta  + 2\pi } \right) = \hat H\left( \theta  \right)$, while the transformed Hamiltonian satisfies the covariant relation: $\tilde H\left( {\theta  + 2\pi } \right) = {{\hat U}_{2\pi }}\tilde H\left( \theta  \right){{\hat U}_{2\pi }}^{ - 1}$.
%
% By choosing a specific gauge, this transformation makes the vector field distributed uniformly.
%
For simplicity, we refer to the former one as ``corner gauge" and to the latter as ``bulk gauge".
Operators under the bulk gauge will be denoted by tilde sign: $h\to \tilde{h}$.
%
% The twist angle (gauge field) will only cause a slight modification in the spectrum, which can be neglected in the thermodynamic limit.
%
By keeping up to the first order of $1/L$ in Eq.~\eqref{eqn:winding_number_z_complex}, we finally obtain the following real-space expression for the MWN (see detailed derivation in Methods):

\begin{equation}
\label{eqn:HOTI_chiral_mean_displacement}
\nu  = \frac{1}{L}\sum\limits_{{E_n} \ne 0} {\langle {\psi _n}|\hat \Gamma \hat {\cal P}|{\psi _n}\rangle },
\end{equation}
where $\hat{\mathcal{P}} = \{\hat{\cal{P}}_{xy}, \hat{\cal{P}}_{xyz}\}$.
We note that there are boundary states localized near each side under both CTBC and OBC.
As mentioned above, these states possess non-zero energy and, consequently, contribute to the computation of MWN.
To elucidate this phenomenon, we evaluate $\langle\psi_n|\hat{\Gamma} \hat{\mathcal{P}}_{xy}|\psi_n\rangle $ for each eigenstate in 2D BBH model, as depicted in Fig.~\ref{fig:FIG_P_bulk_edge}.
Notably, boundary states and bulk states exhibit distinct contributions, manifesting the feature of higher-order topology compared to the conventional topological insulator.

\begin{figure}
\centering
  \includegraphics[width = \columnwidth ]{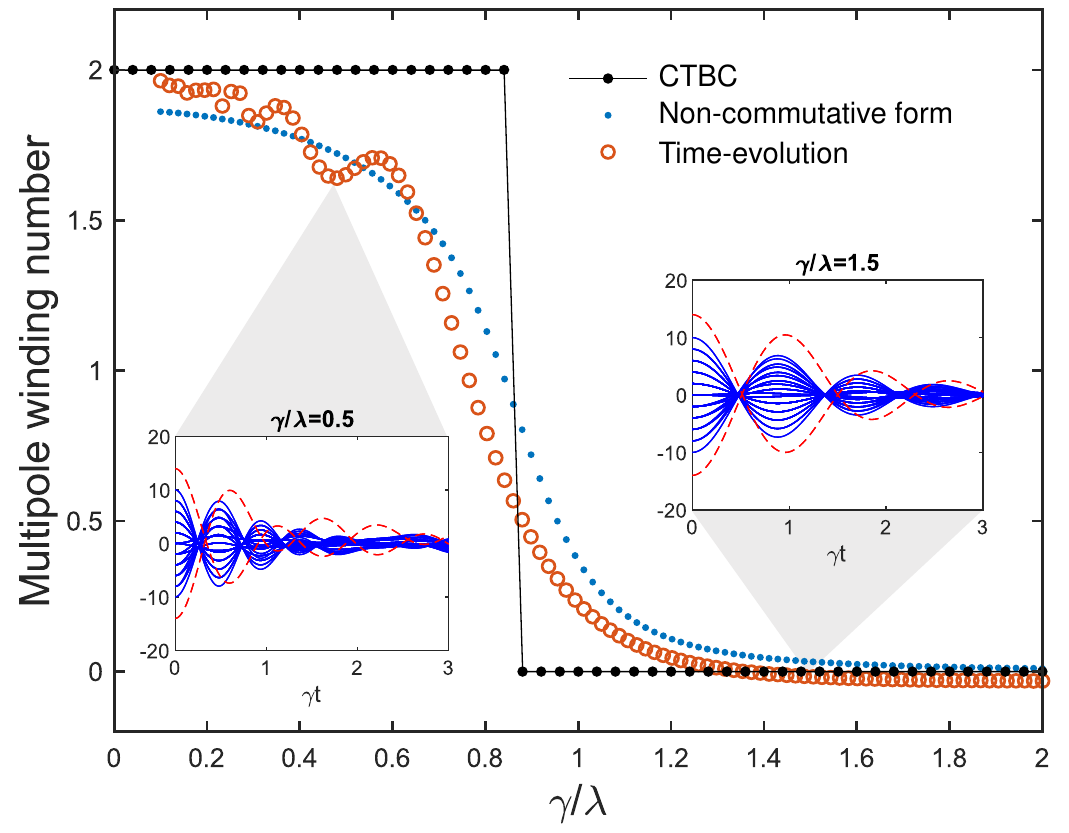} 
  \caption{\label{fig:FIG_Dynamical_WindingNumber}
  \textbf{Numerical simulation of measuring MWN from dynamical evolution.}
  MWN obtained from the CTBC [Eq.~\eqref{eqn:winding_number_z_complex}] (black dots), the non-commutative form [Eq.~\eqref{eqn:HOTI_chiral_mean_displacement}] (blue dots), and the time-evolution [Eq.~\eqref{eqn:nu_bulk_and_edge}] (red circles).
  Bulk initial states are chosen as Fock states ranging from $[L/3, 2L/3] \times [L/3, 2L/3]$.
  For initial states on the edge, we find it is sufficient to use four Fock states localized at the center of the edge: $(1, L/2), (L/2, 1), (L, L/2), (L/2, L)$.
  \emph{Insets}: the time-evolution of $\langle \Psi(t)|\hat{\Gamma}\hat{\cal P}_{xy}|\Psi(t)\rangle$ with
  the red dashed lines for the edge state and the blue lines for the bulk state.
  The system size is chosen as $30\times 30$, and the total evolution time is set as $\gamma t = 3$.
   }
\end{figure}

%\subsection*{Measurement of multipole winding number through dynamical evolution}
%
\textbf{Measurement of multipole winding number through dynamical evolution.}
In practice, one may use dynamical average to obtain the expectation over eigenstates.
Below we show how to measure the MWN for a 2D BBH model.
Under the CTBC, it can be found that there are boundary states localized near the four sides with non-zero energy.
Hence, the MWN should not only relate to the bulk states, but also the boundary states.
%
%These boundary states almost have no overlaps with the bulk states.
%
To attain Eq.~\eqref{eqn:HOTI_chiral_mean_displacement}, it is crucial that the contributions from bulk and boundary states should be both taken into account.
In the BBH model, the numbers of boundary and bulk states are roughly $N_{\mathrm{edge}} = 4L$ and $N_{\mathrm{bulk}} = L^2-4L$.
% with $L$ being the length of the system in each direction.
%
We consider two kinds of initial states: the edge initial state $|{\Psi _{\mathrm{edge}}}\left( 0 \right)\rangle  = \sum\nolimits_{\psi_n \in \mathrm{edge}} {{b_n}|{\psi _n}\rangle }$ and the bulk initial state $|{\Psi _{\mathrm{bulk}}}\left( 0 \right)\rangle  = \sum\nolimits_{\psi_n \in \mathrm{bulk}} {{a_n}|{\psi _n}\rangle }$.
Based upon the time-evolution from these initial states, one can obtain the contribution from the bulk
\begin{eqnarray}
{\nu _{\mathrm{bulk}}} &= &\frac{1}{{\Delta t}}\int_t^{t + \Delta t} {\langle {\Psi _{\mathrm{bulk}}}\left( t \right)|\hat \Gamma {{\hat {\cal P}}_{xy}}|{\Psi _{\mathrm{bulk}}}\left( t \right)\rangle \mathrm{d}t} \nonumber   \\
&\approx& \sum\nolimits_{\psi_n \in \mathrm{bulk}} {|{a_n}{|^2}\langle {\psi _n}|\hat \Gamma {{\hat {\cal P}}_{xy}}|{\psi _n}\rangle }, 
\end{eqnarray}
and the contribution from the edge 
\begin{eqnarray}
{\nu _{\mathrm{edge}}} &=& \frac{1}{{\Delta t}}\int_t^{t + \Delta t} {\langle {\Psi _{\mathrm{edge}}}\left( t \right)|\hat \Gamma {{\hat {\cal P}}_{xy}}|{\Psi _{\mathrm{edge}}}\left( t \right)\rangle \mathrm{d}t} \nonumber  \\
& \approx& \sum\nolimits_{\psi_n \in \mathrm{edge}} {|{b_n}{|^2}\langle {\psi _n}|\hat \Gamma {{\hat {\cal P}}_{xy}}|{\psi _n}\rangle }.
\end{eqnarray}
For simplicity, we assume the initial state is a equal-weight superposition of its corresponding eigenstates, that is $|{a_n}{|^2} \approx \frac{1}{{{L^2} - 4L}}$ and $|{b_n}{|^2} \approx \frac{1}{{4L}}$.
Therefore, the MWN in Eq.~\eqref{eqn:HOTI_chiral_mean_displacement} can be given as
\begin{equation}
\label{eqn:nu_bulk_and_edge}
\nu  \approx \left( {L - 4} \right){\nu _{\mathrm{bulk}}} + 4{\nu _{\mathrm{edge}}}.
\end{equation}
Through numerical simulation, we check the validity of Eq.~\eqref{eqn:nu_bulk_and_edge}.
We simulate the time-evolution from Fock states that localize at the center of the bulk and the edge, respectively.
Then, we calculate $\nu _{\mathrm{bulk}}$ and $\nu _{\mathrm{edge}}$ to obtain the MWN via Eq.~\eqref{eqn:nu_bulk_and_edge}.
These procedures are repeated for different Fock states chosen appropriately to ensure that all eigenstates except for the zero-energy modes are covered.
By averaging them, we obtain the results in Fig.~\ref{fig:FIG_Dynamical_WindingNumber}.
The simulated results are consistent with the ones given by the non-commutative form Eq.~\eqref{eqn:HOTI_chiral_mean_displacement}.
Note that it is difficult to achieve a very long time evolution in realistic experiments. 
In our scheme, it is sufficient to consider the evolution before states reach the corners, which avoids the influence of corner states (see discussions in Sec. D in Supplementary Note 2).

\section*{Discussions} 
In summary, to characterize chiral-symmetric HOTIs, we have introduced the MWN and explored its experimental measurement. 
Through introducing suitable twist operators to transform the gauge field at corners into the bulk, the twist angle serves as a perturbation, leading to the derivation of a real-space formula for the MWN.
This formula provides a practical approach for experimentally probing higher-order topology.
Particularly, our method may be applied to those chiral-symmetric HOTIs with higher winding number in the presence of long-range tunneling \citep{PhysRevLett.131.157201,PhysRevB.108.205135,PhysRevApplied.20.064042}, in which the number of corner states can be hardly distinguished in large systems.
By utilizing the corner twisted boundary condition, we demonstrate the effectiveness of the TBC-like method in studying real-space topological invariants and the bulk-corner correspondence.

In contrast to the conventional TBC, the CTBC exclusively connects corners, leading to the emergence of boundary states near borders in non-trivial cases.
Notably, these boundary states have non-zero energies. 
The MWN is rooted in both bulk and boundary states, setting it apart fundamentally from conventional chiral-symmetric topological insulators.
This distinction necessitates measuring both contributions from the bulk and boundary in experiments.
Furthermore, the appearance of corner states is a consequence of the non-trivial topology of both the bulk and the edge. 
Thus, unlike the conventional topological band theory, the MWN can not be solely characterized with the Bloch states in bulk under the PBC.
This also explains why the MWN can well identify different BOTPs.

In our real-space formulation, the twist operators [Eq.~\eqref{eqn:twist_operator_2D} and Eq.~\eqref{eqn:twist_operator_3D}] significantly differ from the commonly used quadrupole and octupole operators~\citep{PhysRevB.100.245135,PhysRevB.100.245134,PhysRevB.100.245133,PhysRevLett.125.166801,PhysRevB.103.085408}.
This distinction arises from their connection to the topological invariant defined through specific TBC (see detailed derviations in Sec. A in Supplementary Note 2).
Due to the gauge freedom, the twist operator is not unique, allowing for numerous gauge transformations and resulting in various possible real-space formulations using the introduced perturbative expansions. 
Despite these differences, they all share the same physical origin, underscoring the importance of the TBC method in studying real-space formulas for topological invariants.
Moreover, the MWN defined through the CTBC effectively captures information about zero-energy modes. 
This enables the generalization of our method to a broad range of chiral-symmetric HOTIs beyond square or cubic structures, such as the honeycomb lattice~\citep{PhysRevLett.122.086804, PhysRevLett.123.053902,JPSJ.88.104703,PhysRevLett.124.166804},  the kagome lattice \citep{kempkes2019robust, el2019corner,xue2019acoustic,PhysRevB.105.085411,yang2023variable} and even the quasi crystalline lattice \citep{PhysRevLett.123.196401, PhysRevB.100.115311, PhysRevB.102.241102, lv2021realization}.
On the other hand, it is also potential to generalize the MWN to the Floquet HOTIs \citep{PhysRevB.102.094305,PhysRevB.99.045441,PhysRevLett.124.216601,PhysRevLett.124.057001,Ghosh_2024,ghosh2023floquet} with appropriate construction.
The real-space formula for the corresponding MWN can be derived in a similar manner.

\section*{Methods}
%\subsection*{Chiral-symmetric lattice systems and singular value decomposition}
%
\textbf{Chiral-symmetric lattice systems and singular value decomposition.}
Usually, a chiral-symmetric lattice system can be divided into two sublattices obeying a tight-binding Hamiltonian: 
\begin{eqnarray}
    \hat H =  {{\boldsymbol{\psi }}_A}h{\boldsymbol{\psi }}_B^\dag  + {{\boldsymbol{\psi }}_B}{h^\dag }{\boldsymbol{\psi }}_A^\dag ,
\end{eqnarray}
where ${{\boldsymbol{\psi }}_\alpha } \equiv \left( {|{1_\alpha }\rangle ,|{2_\alpha }\rangle , \cdots } \right)$ with $\alpha = \{A, B\}$ indexing the two sublattices and $|i_\alpha\rangle$ being the single-particle position-space basis for the $\alpha$ sublattice.
For convenience, we assume the dimensions of $A$ and $B$ sublattices are equal.
%
%Otherwise, the zero-energy flat band would appear.
%
The Hamiltonian is completely determined by the off-diagonal matrix $h$.
%
% The chiral operator $\hat \Gamma  = \hat{\Gamma}_A  -\hat{\Gamma}_B$, which satisfies $\hat \Gamma \hat H\hat \Gamma  =  - \hat H$ and ${{\hat \Gamma }^{ - 1}} = \hat \Gamma $, is determined by the projector onto the two sublattices $\hat{\Gamma}_\alpha = {\boldsymbol{\psi}}_\alpha {{\boldsymbol{\psi}}_\alpha^\dag}$.
%
The chiral symmetry results in a symmetric spectrum: the eigenvalues of $H$ appear pairwise with opposite values, which are related to the singular values of $h$~\citep{PhysRevB.103.224208}.
The eigen-equation $\hat H|{\Psi _n}\rangle  = {E_n}|{\Psi _n}\rangle $ can be expressed as two coupled equations: $h u_{B}^{[n]} = {E_n}u_{A}^{[n]}$ and ${h^\dag }u_{A}^{[n]} = {E_n}u_{B}^{[n]}$.
Here, the $n$th eigenstate reads as $|{\Psi _n}\rangle  = \frac{1}{{\sqrt 2 }}\sum\nolimits_j {( {u_A^{j,n}|{j_A}\rangle  + u_B^{j,n}|{j_B}\rangle } )} $, and the column vectors $u_{A,B}^{[n]}$ have $u^{j,n}_{A,B}$ as their $j$th elements.
By performing the singular value decomposition: $h = {u_A}\Sigma {u_B}^{ - 1}$, where $\Sigma = \mathrm{diag}(E_1,\cdots,E_n,\cdots)$ is diagonal-positive-definite and $u_{A,B}=[u_{A,B}^{[1]},u_{A,B}^{[2]},\cdots,u_{A,B}^{[L/2]}]$ is an unitary matrix, one can solve the coupled equations.
It can be seen that the singular values of matrix $h$ are exactly half of the energy spectrum.

%\subsection*{Derivation of the real-space formula of multipole winding number}
%
\textbf{Derivation of the real-space formula of multipole winding number.}
Choosing $r=1$, under the bulk gauge, the MWN can still be expressed as Eq.~\eqref{eqn:winding_number_z_complex}.
Keeping the terms up to the first order of $1/L$, we have $\tilde{h}\left( \theta  \right) = h \left( 0 \right) + \frac{\theta }{L}{[ {\frac{\partial }{{\partial \theta /L}}\tilde{h}\left( \theta  \right)} ]_{\theta  = 0}} + O\left( {\frac{1}{{{L^2}}}} \right)$ and ${\rm{Tr}}[ {h{{\left( {r,\theta } \right)}^{ - 1}}{\partial _\theta }h\left( {r,\theta } \right)} ] \approx {\rm{Tr}}\{ {{h^{ - 1}}\left( 0 \right){{[ {{{\partial_\theta }}\tilde h\left( \theta  \right)} ]}_{\theta  = 0}}} \}$ being a constant.
Therefore, the MWN can be given as $ \nu \approx - i{\rm{Tr}}\{ {{h^{ - 1}}\left( 0 \right){{[ {{{\partial_\theta }}\tilde{h} \left( \theta  \right)} ]}_{\theta  = 0}}} \}$.
%
% However, we still need to compute the partial derivative of $\tilde{h} \left( \theta  \right)$.
%
Using $\tilde h\left( \theta  \right) = {\boldsymbol{\psi }}_A^\dag \tilde H\left( \theta  \right){{\boldsymbol{\psi }}_B}$, we have $\tilde h = {\cal U}_{2\pi }^A\tilde h\left( 0 \right){( {{\cal U}_{2\pi }^B} )^{ - 1}}$ with ${{\hat U}_{2\pi }} = \exp ( {i2\pi \hat{\mathcal{P}}/L} )$, where $\hat{\mathcal{P}} = \{\hat{\cal{P}}_{xy}, \hat{\cal{P}}_{xyz}\}$ are both diagonal in the position basis.
For simplicity, we define the projected twist operator matrices in the sector $\alpha  = \{A, B\}$: ${\cal U}_{2\pi }^\alpha  \equiv {\boldsymbol{\psi }}_\alpha ^\dag {{\hat U}_{2\pi }}{{\boldsymbol{\psi }}_\alpha }$.
To ensure the gauge invariance, we impose the constraint $\det \tilde h\left( {2\pi } \right) = \det \tilde h\left( 0 \right)$.
Thus, we require ${\cal U}_{2\pi }\equiv {\cal U}_{2\pi }^A = {\cal U}_{2\pi }^B$, which can be easily satisfied by setting an identical position for all sublattices within a unit cell.
Up to the first order of $1/L$, we have
${\left[ {{\partial _\theta }\tilde h\left( \theta  \right)} \right]_{\theta  = 0}} \approx \frac{{\tilde h\left( {2\pi } \right) - \tilde h\left( 0 \right)}}{{2\pi }} = \frac{{{\cal U}_{2\pi }h{{ {{\cal U}_{2\pi }^{ - 1}} }} - h}}{{2\pi }}$.
%
%where we omit the dependence on the twist angle $\theta$ when $\theta = 0$ for brevity.
%
To gain higher accuracy, one may consider a higher-order finite difference (see
details in Sec. B of Supplementary Note 2).
Then, a rather simpler real-space formula can be obtained: $\nu  = \frac{1}{{2\pi i}}{\rm{Tr}}\left( {{h^{ - 1}}{\cal U}_{2\pi }h{{ {{\cal U}_{2\pi }^{ - 1}} }} - I} \right)$.
This leads to two equivalent real-space formulas: (i) the {Bott index form} and (ii) the {non-commutative form}.

Since ${{h^{ - 1}}{\cal U}_{2\pi }h{{ {{\cal U}_{2\pi }^{ - 1}} }}}$ is close to an identity matrix, in the thermodynamic limit $L\to \infty$, one can obtain the {Bott index form} $\nu  = \frac{1}{{2\pi i}}{\rm{Tr}}\left[ {\log \left( {{h^{ - 1}}{\cal U}_{2\pi }h{{ {{\cal U}_{2\pi }^{ - 1}} }}} \right)} \right]$.
This formula is analogous to the multipole chiral number~\citep{PhysRevLett.128.127601}, which also takes the Bott index form.
The multipole chiral number is defined based on the quadrupole and octupole of polarization.
It can be proved that the multipole chiral number is equivalent to the real-space winding number defined through a modified CTBC  (see details in Sec. A of Supplementary Note 2)..
The modified CTBC still connects the corners, while only those tunnelings related to one of the corners are threaded by a twist angle.
In this scenario, the modified MWN captures the number of zero-energy modes induced in one specific corner.
Therefore, although the generators of the twist operators [Eq.~\eqref{eqn:twist_operator_2D} and Eq.~\eqref{eqn:twist_operator_3D}] are quite different from the quadrupole and octupole, they share the same physical origin.

Using the Baker-Campbell-Hausdorff (BCH) formula,
we have ${{\cal U}_{2\pi }}h{{\cal U}_{2\pi }^{ - 1}} \approx h + 2\pi i\left[ {P,h} \right]/L$ with  $P = {\boldsymbol{\psi }}_A^\dag \hat{\mathcal{P}}  {\boldsymbol{\psi }}_A$ denoting the generator projected onto the sector $A$, and then obtain the {non-commutative form} $\nu  = \frac{1}{L}\mathrm{Tr}\left( {h^{ - 1}}\left[ {{{P}},h} \right] \right)$.
However, this approximation only holds for systems with OBC and $L \to \infty$.
In a finite system with CTBC, this approximation is poor because it always evaluates to zero if $h$ is invertible: ${\rm{Tr}}\left( {{h^{ - 1}}\left[ {P,h} \right]} \right) = {\rm{Tr}}\left( {{h^{ - 1}}Ph - P} \right) = 0$.
This issue arises from the slow convergence of the BCH formula and the omission of higher-order terms (see
discussions in Sec. B of Supplementary Note 2).
Meanwhile, as the CTBC connects the corners, general definition of the position near the corner become ill-defined, which also leads to the poor performance of the first-order expansion.
However, under the OBC, the BCH formula converges fast because position-dependent operators are well-defined.
In the presence of zero modes, $h$ becomes singular.
In this case, we need to rule out the zero singular values and the associated vectors and use the pseudo-inverse formulation: $h^{-1}\approx \tilde{u}_B \tilde{\Sigma} ^{-1} \tilde{u}_A^{\dag}$ and then the non-commutative form will be non-zero.
%%

%%%
The above forms for MWN are not straightforward to be measured in experiments.
Below we derive a formula suitable for experimental measurement.
Firstly, the flattened Hamiltonian can be written as $\hat Q = {{\hat{\mathbb{P}}}_ + } - {{\hat{\mathbb{P}}}_ - }$ with ${{\hat{\mathbb{P}}}_ - } = \sum\nolimits_{{E_n} < 0} {|{\psi _n}\rangle \langle {\psi _n}|} $ and ${{\hat{\mathbb{P}}}_ + } = \sum\nolimits_{{E_n} > 0} {|{\psi _n}\rangle \langle {\psi _n}|} $, where $E_n$ is the eigenenergy for the eigenstate $|\psi_n\rangle $.
The flattened Hamiltonian shares the same eigenstates of the original Hamiltonian and the zero-energy mode is naturally excluded.
%
%Due to the chiral symmetry, we have ${{\hat{\mathbb{P}}}_ + } =\hat \Gamma {{\hat{\mathbb{P}}}_ - }  \hat \Gamma $.
%
Replacing the off-diagonal matrix $h$ with the flattened matrix $q\equiv {\boldsymbol{\psi }}_A^\dag \hat Q{{\boldsymbol{\psi }}_B}$ and  $q^{-1}\equiv {\boldsymbol{\psi }}_B^\dag \hat Q{{\boldsymbol{\psi }}_A}$, we have $\nu  = \frac{1}{L}\mathrm{Tr}\left( {q^{ - 1}}\left[ {{{P}},q} \right] \right)$.
Therefore, we obtain a real-space formula (see detailed derivations Sec. C of Supplementary Note 2),
\begin{equation}
\label{eqn:HOTI_chiral_mean_displacement}
\nu  = \frac{1}{L}\sum\limits_{{E_n} \ne 0} {\langle {\psi _n}|\hat \Gamma \hat {\cal P}|{\psi _n}\rangle } ,
\end{equation}
which works well under OBC.
With Eq.~\eqref{eqn:HOTI_chiral_mean_displacement}, we can extract the MWN by measuring the expectations of  $\hat{\Gamma}\hat{\cal P}_{xy}$ for 2D systems or $\hat{\Gamma}\hat{\cal P}_{xyz}$ for 3D systems.

\section*{Acknowledgments}
We acknowledge useful discussions with Yonguan Ke and Wenjie Liu. This work is supported by the National Key Research and Development Program of China (Grant No. 2022YFA1404104), and the National Natural Science Foundation of China (Grant No. 12025509 and Grant No. 12247134).

%aipnum4-2.bst 2019-01-14 (MD) hand-edited version of apsrev4-1.bst

\end{document}

\title{Supplemental Information for ``Probing Chiral-Symmetric Higher-Order Topological Insulators with Multipole Winding Number"}
% Real-space characterization of winding number from the bulk in chiral-symmetric higher-order topological insulators
% (order of authors and affiliations are tentative) 

\author{Ling Lin$^{1,2,3}$}
\author{Chaohong Lee$^{1,2,3}$}
\email{Corresponding author. Email: chleecn@szu.edu.cn, chleecn@gmail.com}

\affiliation{$^{1}$Institute of Quantum Precision Measurement, State Key Laboratory of Radio Frequency Heterogeneous Integration, Shenzhen University, Shenzhen 518060, China}
\affiliation{$^{2}$College of Physics and Optoelectronic Engineering, Shenzhen University, Shenzhen 518060, China}
\affiliation{$^{3}$Quantum Science Center of Guangdong-Hongkong-Macao Greater Bay Area (Guangdong), Shenzhen 518045, China}

\date{\today}

%=========Abstract=========
% No abstract
% ========Main body==========

\maketitle
\tableofcontents  % optional

\makeatletter
\renewcommand{\theequation}{S\arabic{equation}}
\renewcommand{\thefigure}{S\arabic{figure}}
\renewcommand{\thesection}{S\arabic{section}}
\section{Supplementary Note 1: Multipole winding number and the corner twisted boundary condition}
\subsection{Quantization of higher-order polarizations and its relation to the multipole winding number}
With the CTBC, one can define a multipole polarization for eigenstates below $E_n=0$ via the non-Abelian Berry phase formalism:
\begin{equation}
\label{eqn:higher_order_polarization}
p = - \frac{1}{{2\pi }}\int_0^{2\pi } {{\rm{d}}\theta \;{\rm{Tr}}\left( {{\boldsymbol{\Psi }}_\theta ^\dag i{\partial _\theta }{{\boldsymbol{\Psi }}_\theta }} \right)} 
\end{equation}
where ${{\boldsymbol{\Psi }}_\theta } = \left( {|{\Psi _1}\left( \theta  \right)\rangle , \cdots ,|{\Psi _n}\left( \theta  \right)\rangle } \right),\;{E_n} < 0$, and $|{\Psi _n}\left( \theta  \right)\rangle $ is the $n$th eigenstate.
Due to the chiral symmetry, the polarization of the positive-energy sector is the same since $\Gamma |{\Psi _n}\rangle  = |{\Psi _{ - n}}\rangle $, with $ |{\Psi _{ - n}}\rangle$ being the eigenstate with opposite eigenenergy.
This multipole polarization is related to quantized bulk charge pump and satisfies the bulk-edge correspondence \citep{PhysRevLett.128.246602}.
Note that Eq.~\eqref{eqn:higher_order_polarization} is only gauge-invariant modulo $1$.
Here, we adopt an idea in Ref.~\citep{PhysRevLett.113.046802} to prove that the higher-order polarization in Eq.~\eqref{eqn:higher_order_polarization} is related to the MWN: $p=\nu/2\mod 1$.
Firstly, with the flattened Hamiltonian at $r=1$, the MWN in can be equivalently written as 
\begin{equation}
\nu  = \frac{1}{{2\pi i}}\int_0^{2\pi } {{\rm{d}}\theta \;{\rm{Tr}}\left[ {q{{\left( {\theta } \right)}^{ - 1}}{\partial _\theta }q\left( {\theta } \right)} \right]} .
\end{equation}
Using $q\left( \theta  \right) = {u_A}\left( \theta  \right)u_B^{ - 1}\left( \theta  \right)$ according to the singular value decomposition (SVD), we have
\begin{eqnarray}
\label{eqn:appendix_nu_nua_nub}
\nu &=& \frac{1}{{2\pi i}}\int_0^{2\pi } {{\rm{d}}\theta \;{\rm{Tr}}\left[ {q{{\left( \theta \right)}^{ - 1}}{\partial _\theta }q\left( \theta \right)} \right]} \nonumber \\
 &=& \frac{1}{{2\pi i}}\int_0^{2\pi } {{\rm{d}}\theta \;{\rm{Tr}}\left\{ {\left[ {{u_B}\left( \theta \right)u_A^{ - 1}\left( \theta \right)} \right]{\partial _\theta }\left[ {{u_A}\left( \theta \right)u_B^{ - 1}\left( \theta \right)} \right]} \right\}} \nonumber \\
 &=& \frac{1}{{2\pi i}}\int_0^{2\pi } {{\rm{d}}\theta \;{\rm{Tr}}\left\{ {u_A^{ - 1}\left( \theta \right){\partial _\theta }{u_A}\left( \theta \right)} \right\}}  + \frac{1}{{2\pi i}}\int_0^{2\pi } {{\rm{d}}\theta \;{\rm{Tr}}\left\{ {{u_B}\left( \theta \right){\partial _\theta }u_B^{ - 1}\left( \theta \right)} \right\}} \nonumber \\
 &= &\frac{1}{{2\pi i}}\int_0^{2\pi } {{\rm{d}}\theta \;{\rm{Tr}}\left\{ {u_A^{ - 1}\left( \theta \right){\partial _\theta }{u_A}\left( \theta \right)} \right\}}  - \frac{1}{{2\pi i}}\int_0^{2\pi } {{\rm{d}}\theta \;{\rm{Tr}}\left\{ {u_B^{ - 1}\left( \theta \right){\partial _\theta }{u_B}\left( \theta \right)} \right\}} \nonumber \\
 &=& {\nu _A} - {\nu _B},
\end{eqnarray}
where we denote the two winding numbers respectively corresponding to the $A$ and $B$ sublattices:
\begin{eqnarray}
{\nu _A} &\equiv& \frac{1}{{2\pi i}}\int_0^{2\pi } {{\rm{d}}\theta \;{\rm{Tr}}\left\{ {u_A^{ - 1}\left( \theta  \right){\partial _\theta }{u_A}\left( \theta  \right)} \right\}} \in \mathbb{Z} \nonumber  \\
{\nu _B} &\equiv& \frac{1}{{2\pi i}}\int_0^{2\pi } {{\rm{d}}\theta \;{\rm{Tr}}\left\{ {u_B^{ - 1}\left( \theta  \right){\partial _\theta }{u_B}\left( \theta  \right)} \right\}}  \in \mathbb{Z},
\end{eqnarray}
since ${u_{A,B}}\left( \theta  \right)$ are \emph{unitary matrices}.
We would like to point out that the matrices ${u_{A,B}}\left( \theta  \right)$ are not gauge-invariant, and therefore there is gauge freedom in these two winding numbers.
%
% Thus, we should impose that ${u_{A,B}}\left( \theta  + 2\pi \right) = {u_{A,B}}\left( \theta  \right)$ for completeness.
%
However, $q\left( \theta  \right) = {u_A}\left( \theta  \right)u_B^{ - 1}\left( \theta  \right)$ and its associated MWN is gauge-invariant.
Thus, the difference between $\nu_A$ and $\nu_B$ is also gauge-invariant
Then, using the notation: ${{\boldsymbol{ \psi }}_\alpha } \equiv \left( {|{1_\alpha }\rangle ,|{2_\alpha }\rangle , \cdots } \right)$ with $\alpha = A, B$, we write the eigenstate as:
\begin{equation}
|{\Psi _n}\rangle  = \frac{1}{{\sqrt 2 }}\left( {{{\boldsymbol{\psi}}_A}u_{A}^{[n]} - {{\boldsymbol{\psi}}_B}u_{B}^{[n]}} \right),
\end{equation}
and therefore
\begin{eqnarray}
{{\boldsymbol{\Psi }}_\theta }& =& \left( {|{\Psi _1}\left( \theta  \right)\rangle , \cdots ,|{\Psi _n}\left( \theta  \right)\rangle } \right) \nonumber \\
&=& \frac{1}{{\sqrt 2 }}\left[ {{{\boldsymbol{\psi}}_A}{u_A}\left( \theta  \right) - {{\boldsymbol{\psi}}_B}{u_B}\left( \theta  \right)} \right].
\end{eqnarray}
Substitute it into Eq.~\eqref{eqn:higher_order_polarization} yields the following relation:
\begin{eqnarray}
p &=&  - \frac{1}{{2\pi }}\int_0^{2\pi } {{\rm{d}}\theta \;{\rm{Tr}}\left( {{\boldsymbol{\Psi }}_\theta ^\dag i{\partial _\theta }{{\boldsymbol{\Psi }}_\theta }} \right)} \nonumber\\
 &=& \frac{1}{{4\pi i}}\int_0^{2\pi } {{\rm{d}}\theta \;{\rm{Tr}}\left( {\left[ {{u_A}^{ - 1}\left( \theta  \right){\boldsymbol{\psi }}_A^\dag  - {u_B}^{ - 1}\left( \theta  \right){\boldsymbol{\psi }}_B^\dag } \right]{\partial _\theta }\left[ {{{\boldsymbol{\psi }}_A}{u_A}\left( \theta  \right) - {{\boldsymbol{\psi }}_B}{u_B}\left( \theta  \right)} \right]} \right)} \nonumber\\
&=& \frac{1}{{4\pi i}}\int_0^{2\pi } {{\rm{d}}\theta \;{\rm{Tr}}\left[ {\left( {{u_A}^{ - 1}\left( \theta  \right){\boldsymbol{\psi }}_A^\dag } \right){\partial _\theta }\left( {{{\boldsymbol{\psi }}_A}{u_A}\left( \theta  \right)} \right)} \right]}  \nonumber\\
&& + \frac{1}{{4\pi i}}\int_0^{2\pi } {{\rm{d}}\theta \;{\rm{Tr}}\left[ {\left( {{u_B}^{ - 1}\left( \theta  \right){\boldsymbol{\psi }}_B^\dag } \right){\partial _\theta }\left( {{{\boldsymbol{\psi }}_B}{u_B}\left( \theta  \right)} \right)} \right]} \nonumber\\
 &=& \frac{1}{{4\pi i}}\int_0^{2\pi } {{\rm{d}}\theta \;{\rm{Tr}}\left[ {{u_A}^{ - 1}\left( \theta  \right){\partial _\theta }{u_A}\left( \theta  \right)} \right]}  + \frac{1}{{4\pi i}}\int_0^{2\pi } {{\rm{d}}\theta \;{\rm{Tr}}\left[ {{u_B}^{ - 1}\left( \theta  \right){\partial _\theta }{u_B}\left( \theta  \right)} \right]} \nonumber\\
&=& \frac{1}{2}\left( {{\nu _A} + {\nu _B}} \right),
\end{eqnarray}
where we have used the orthogonal relation: ${\boldsymbol{\psi }}_A^\dag {{\boldsymbol{\psi }}_B} = 0$ and ${\boldsymbol{\psi }}_{A,B}^\dag {{\boldsymbol{\psi }}_{A,B}} = 1$.
Combining with Eq.~\eqref{eqn:appendix_nu_nua_nub}, there are following relations:
\begin{eqnarray}
2p + \nu  &=& 2{\nu _A} \in 2 \mathbb{Z}\nonumber \\
2p - \nu  &=& 2{\nu _B} \in 2 \mathbb{Z}.
\end{eqnarray}
The above equations infer that $2p$ and $\nu$ should have the same parity.
Since the polarization is only gauge-invariant modulo 1, we can therefore conclude that $p=\nu/2 \mod 1$.
This explains why the multipole polarization is found to maintain quantization in the presence of disorders provided that the chiral symmetry is preserved. 

\subsection{Finite-size effect to the Multipole winding number}
As mentioned in the main text, the MWN measures the number of singular points in $h(r,\theta)$ encircled by $r=1$.
The bulk-corner correspondence of the MWN is for the CTBC.
Under the OBC, corner states are localized at at the corner and usually have zero energy.
However, in a finite system, the wavefunction of the corner state may have overlap with each other and then hybridize.
This phenomenon is particularly obvious near the phase transition point, where the localization length of the corner state become considerably long.
Thus, the energy of these corner states would deviate from zero, causing the change of the MWN even if the system is not at the phase transition point under PBC.

\begin{figure}
\centering
  \includegraphics[width = \columnwidth ]{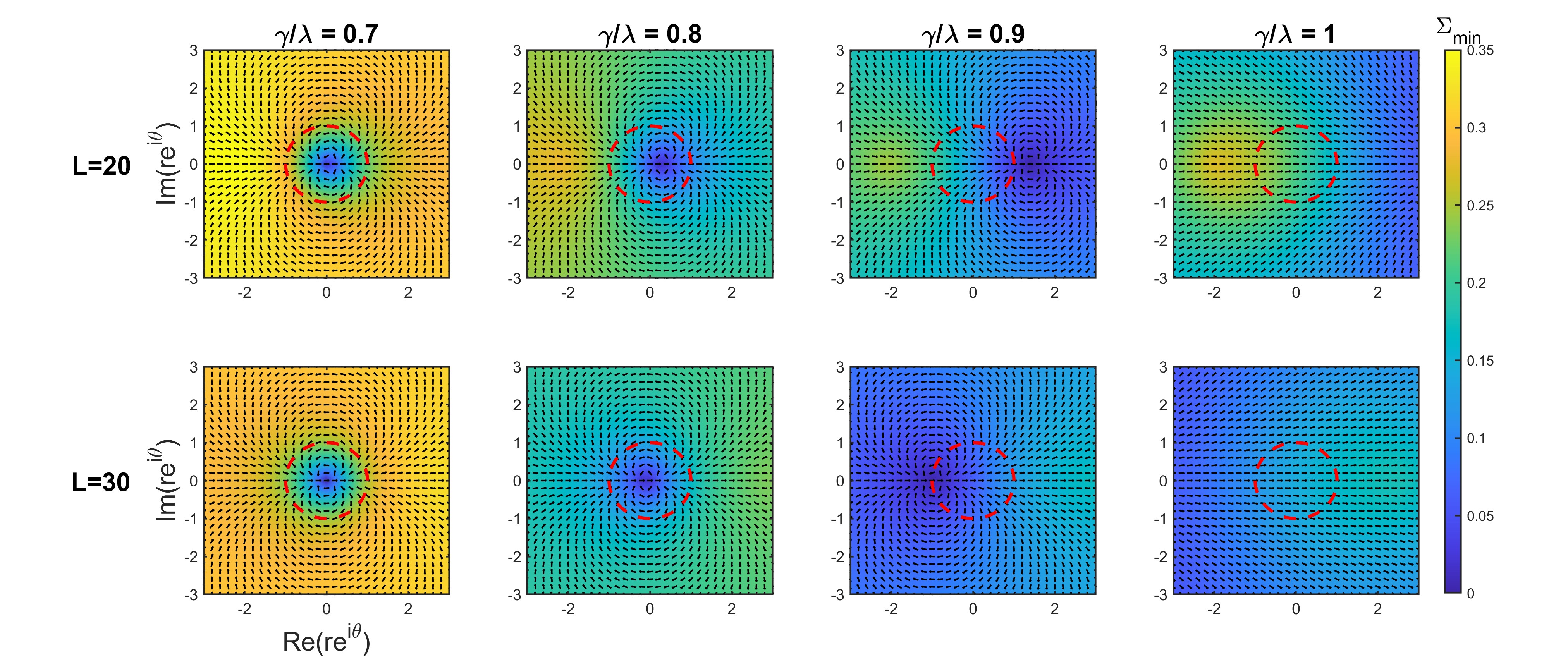}
  \caption{\label{fig_Appendix:FIG_Appendix_r_theta_transition}
  Diagram of $\det h(r,\theta)$ for different parameters near the phase transition point of 2D BBH model.
  The red dashed circle hints the integral loop $\mathcal{C}$ ($r=1$).
    }
\end{figure}

\begin{figure}
\centering
  \includegraphics[width = 1\columnwidth ]{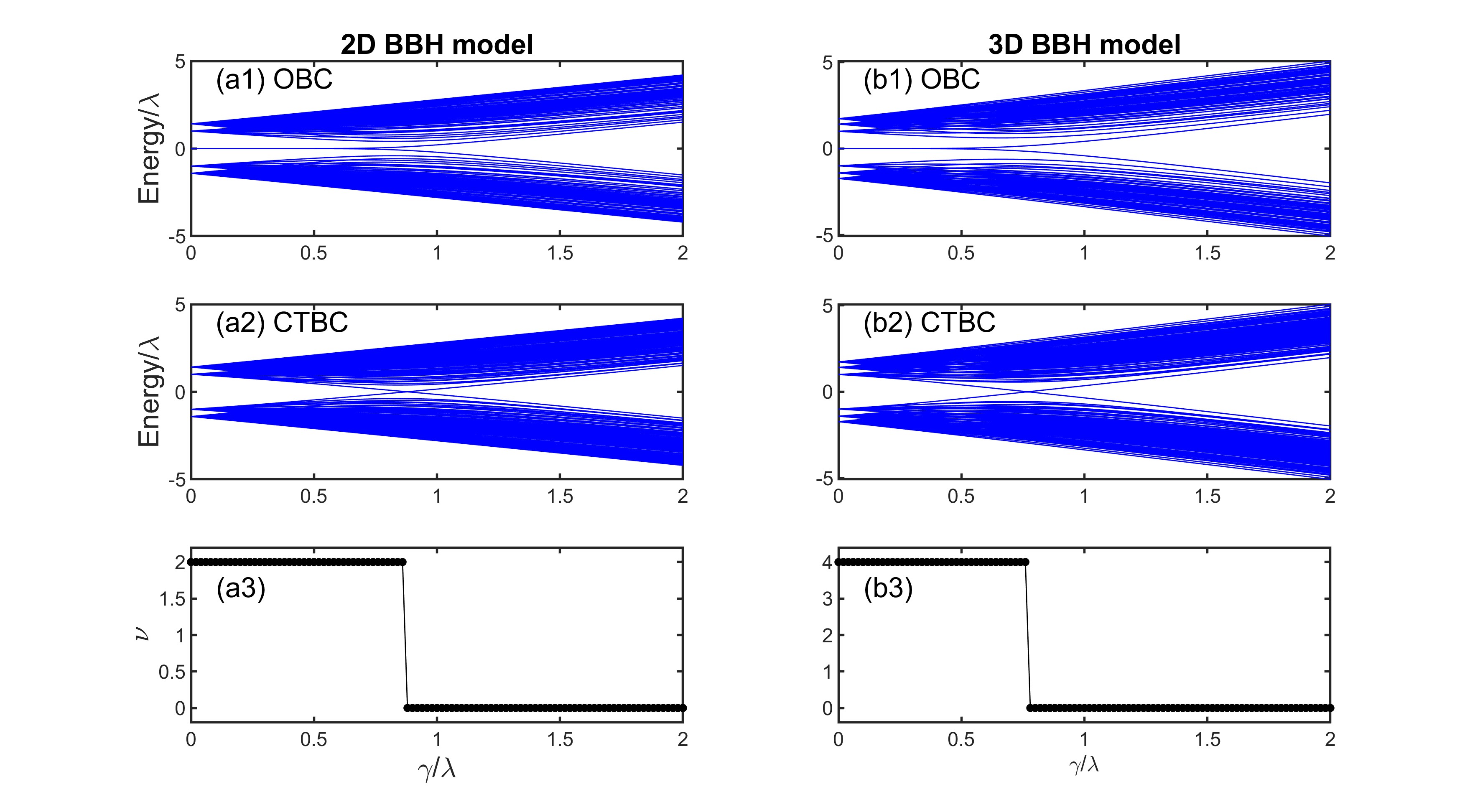}
  \caption{\label{fig_Appendix:FIG_Appendix_OBC_CTBC_comparison}
  Comparison of the spectrum  between the OBC and the CTBC (with $r=1$ and $\theta = 0$).
  (a1-a3) are computed from the 2D BBH model with $L=20$.
  The zero-energy modes are 4-fold degenerate in (b1) 
  (b1-b3) are computed from the 3D BBH model with $L=12$.
  The zero-energy modes are 8-fold degenerate in (b1) 
  The MWN is computed through the CTBC.
  }
\end{figure}

Here we numerically demonstrate how the MWN changes near the phase transition point.
We compute the diagram of $\det h(r,\theta)$ under CTBC in a 2D BBH model for various parameters, see Fig.~\ref{fig_Appendix:FIG_Appendix_r_theta_transition}.
It can be seen that the singular point is shifted in the complex plane near the phase transition point.
Specifically, one can see that at $\gamma/\lambda = 0.9 < 1$ in the $L=20$ system, the singular point has been shifted out of the circle $r=1$, meaning that the MWN is zero at this point.
This is in contrast to the phase transition point under PBC due to the finite-size effect.
It can be also noted that the singular point is shifted along $\theta=0$ for odd number of cells ($N_{cell}\in 2\mathbb{Z}-1$).
For even number of cells ($N_{cell}\in 2\mathbb{Z}$), it is shifted along $\theta = \pi$.
In the thermodynamic limit, the phase transition point determined by the MWN is expected to be consistent with the phase transition point $|\gamma/\lambda| = 1$ obtained from the multipole polarization method under the PBC, as already shown in the Fig.~2 of the main text.
For further comparison, we have computed the spectrum under OBC and CTBC in 2D and 3D BBH model, see Fig.~\ref{fig_Appendix:FIG_Appendix_OBC_CTBC_comparison}.
In both OBC and CTBC, we can identify that the phase transition points significantly deviate from $\gamma/\lambda = 1$ due to the finite size effect.
The two spectrum are quite similar.
The gap-closing point under the CTBC and the phase transition point given by the MWN both successfully predict the disappearance of the zero-energy modes under the OBC.
Thus, it can be seen that the MWN defined through CTBC can well capture the topological nature of the chiral HOTI for finite systems.

\subsection{Distribution of gauge field under uniform gauge in corner twisted boundary condition}
\begin{figure}
\centering
  \includegraphics[width = 0.7\columnwidth ]{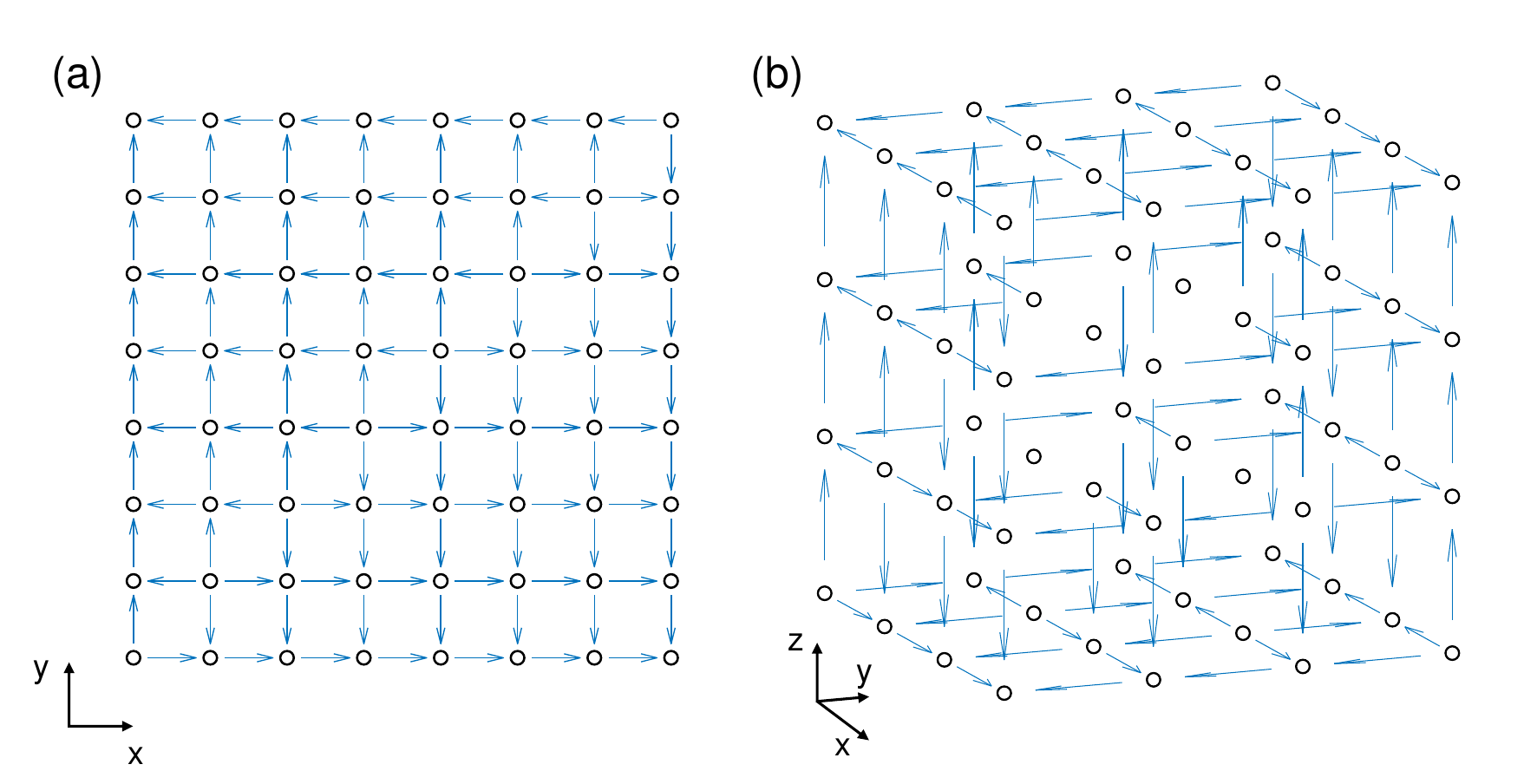}
  \caption{\label{fig_Appendix:FIG_Appendix_Gauge_field_uniform_gauge}
  Distribution of the gauge field inside the bulk of the (a) 2D square lattice and (b) 3D cubic lattice under the uniform gauge.
  Arrows imply the direction of the gauge field between two adjacent sites and the strength is all equal to $\theta/L$.
  For simplicity, gauge fields at corners are not presented here.
  }
\end{figure}

By using the twist operators $\hat{U}_\theta = \{\hat{U}_\theta^{2D}, \hat{U}_\theta^{3D} \}$ introduced in the main text, we can transform the system from the boundary gauge to the bulk gauge through the unitary transformation:
\begin{equation}
    \tilde H\left( \theta  \right) = {\hat U_\theta }\hat H\left( \theta  \right){\hat U_\theta }^{ - 1}.
\end{equation}
As illustrated in the main text, the gauge fields under the boundary gauge only distributes at the corner, which only affects tunnelings between corners.
Under the bulk gauge, the gauge field is not only distributed at the corners but also spread to the bulk, as schematically depicted in Fig.~\ref{fig_Appendix:FIG_Appendix_Gauge_field_uniform_gauge}.
The strength of the gauge field is ``dilluted" to $\theta/L$ at each bond and therefore can be well treated as a perturbation for sufficiently large system.
Note that if the geometry of the lattice is not square or cubic, it is still possible to revise the twist operator and find a suitable gauge to make the gauge field a perturbation.

\section{Supplementary Note 2: Real-space formula of the multipole winding number and real-space measurement}
\subsection{Multipole chiral number as a winding number defined through modified corner twisted boundary condition}
In this section, we show that the multipole chiral number for chiral-symmetric HOTI proposed in Ref.~\citep{PhysRevLett.128.127601} can be derived from the winding number defined through \emph{modified} CTBC.
In many works, the quadrupole operator in 2D or octupole operator in 3D are frequently used to investigate the topological property of HOTIs \citep{PhysRevB.100.245135,PhysRevB.100.245134,PhysRevB.100.245133,PhysRevLett.125.166801,PhysRevB.103.085408}:
\begin{eqnarray}
{{\hat Q}_{xy}} &=& -\sum\limits_{x,y} {xy{{\hat n}_{x,y}}} \nonumber \\
{{\hat O}_{xyz}} &=& \sum\limits_{x,y,z} {xyz{{\hat n}_{x,y,z}}} ,
\end{eqnarray}
where the minus sign in ${{\hat Q}_{xy}} $ is for convenience.
They constitute the multipole twist operators for 2D and 3D respectively
\begin{eqnarray}
\label{eqn:Appendix_twist_operators_Uxyz}
\hat U_\theta ^Q &=& \exp \left( {i\frac{{2\pi }}{{{L^2}}}{{\hat Q}_{xy}}} \right)  \nonumber \\
\hat U_\theta ^O &=& \exp \left( {i\frac{{2\pi }}{{{L^3}}}{{\hat O}_{xyz}}} \right).
\end{eqnarray}
Without loss of generality, we shall still consider the equal length in each dimension.
The multipole chiral numbers for 2D and 3D system are defined through \citep{PhysRevLett.128.127601}
\begin{eqnarray}
\label{eqn_appendix:nu_xy_and_xyz}
 {\nu _{xy}} &=& \frac{1}{{2\pi i}}{\rm{Tr}}\left\{ {\log \left[ {\bar Q_{xy}^A{{\left( {\bar Q_{xy}^B} \right)}^\dag }} \right]} \right\}, \\
 {\nu _{xyz}} &=& \frac{1}{{2\pi i}}{\rm{Tr}}\left\{ {\log \left[ {\bar O_{xyz}^A{{\left( {\bar O_{xyz}^B} \right)}^\dag }} \right]} \right\},
\end{eqnarray}
where
\begin{eqnarray}
\bar Q_{xy}^{A,B} = {u'}_{A,B}^{ - 1}{\cal U}_{2\pi }^{xy}u{'_{A,B}} \\
\bar O_{xyz}^{A,B} = {u'}_{A,B}^{ - 1}{\cal U}_{2\pi }^{xyz}u{'_{A,B}}
\end{eqnarray}
with twist operator projected to the $A$ sublattices: ${\cal U}_{2\pi }^{xy} = {\boldsymbol{\psi }}_A^\dag \hat U_{2\pi }^Q{{\boldsymbol{\psi }}_A},\;{\cal U}_{2\pi }^{xyz} = {\boldsymbol{\psi }}_A^\dag \hat U_{2\pi }^O{{\boldsymbol{\psi }}_A}$.
Again, we have assumed that the twist operator has the same form in the two sublattices lattice: ${\boldsymbol{\psi }}_A^\dag \hat U_{2\pi }^Q{{\boldsymbol{\psi }}_A} = {\boldsymbol{\psi }}_B^\dag \hat U_{2\pi }^Q{{\boldsymbol{\psi }}_B} $ and ${\boldsymbol{\psi }}_A^\dag \hat U_{2\pi }^O{{\boldsymbol{\psi }}_A} = {\boldsymbol{\psi }}_B^\dag \hat U_{2\pi }^Q{{\boldsymbol{\psi }}_B} $, which can be achieved by requiring the definitions of each sublattice's position inside the cell are the same.
Otherwise, the result is affected by the definition of positions.
The multipole chiral number in Eq.~\eqref{eqn_appendix:nu_xy_and_xyz} has been proved capable to characterize different topological phases in chiral-symmetric HOTI even in the presence of disorders that preserve the chiral symmetry.
Specifically, the relations between multipole chiral numbers and the number of zero-energy states in 2D square and 3D cubic systems with open boundary are
\begin{eqnarray*}
N_{corner}^{2D} = 4 \nu_{xy}, \\
N_{corner}^{3D} = 8 \nu_{xy}.
\end{eqnarray*}
This result is in contrast to the MWN defined in the main text of our work, and we will explain it later.

\begin{figure}[htp]
\centering
  \includegraphics[width = 0.8\columnwidth ]{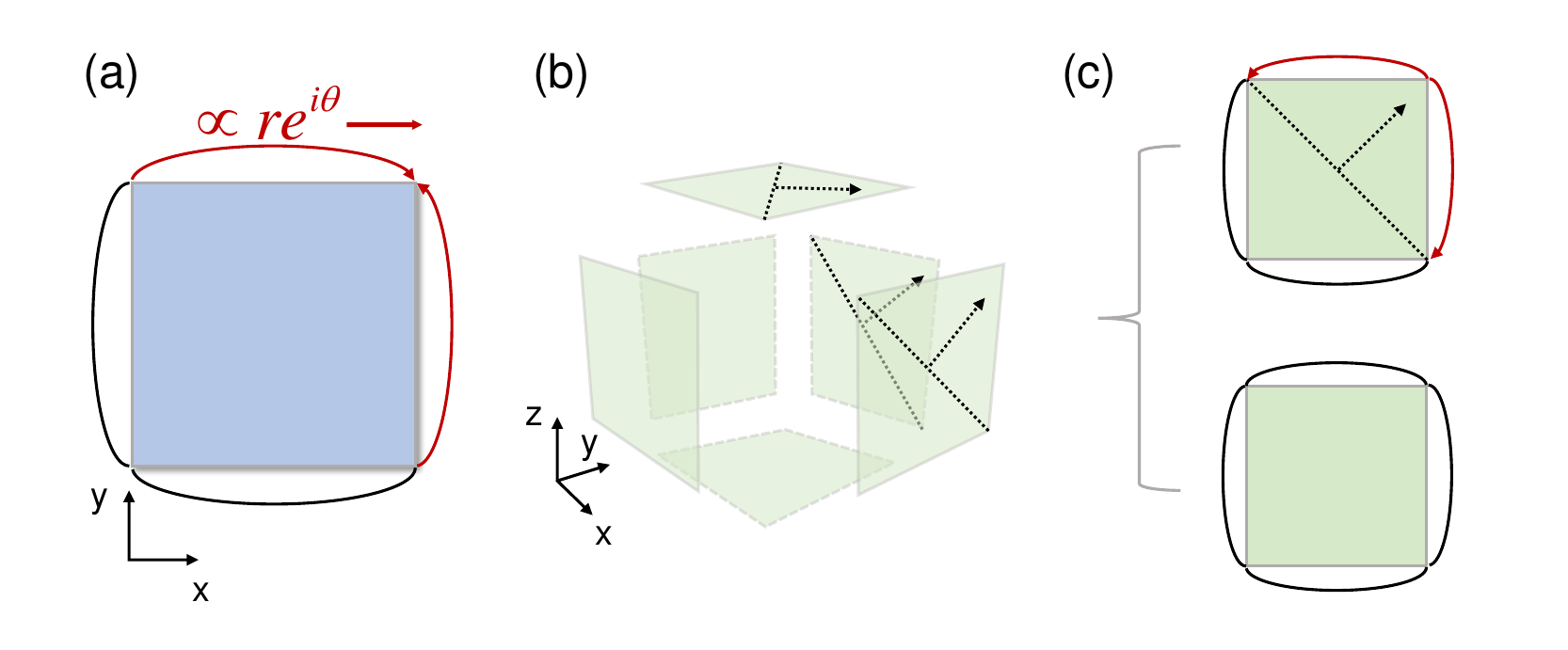}
  \caption{\label{fig_Appendix:FIG_Appendix_Half_CTBC_illustration}
  Schematic illustration of the reduced corner TBC for 2D HOTI and 3D HOTI.
  (a) depicts the 2D case. 
  Corners are still connected, but only the tunneling involving the top-right corner are addressed by twist angle.
  (b) depicts the 3D case with only one corners are addressed with the twist angle (gauge field), while the others are not.
  Specifically, three of the faces indicated by dashed lines and arrows are threaded by gauge field (twist angle) on one of the corners, see (c).
  In other faces without the dashed indicator, the four corners are simply connected (implied by black lines).
  }
\end{figure}

\begin{figure}[htp]
\centering
  \includegraphics[width = 0.7\columnwidth ]{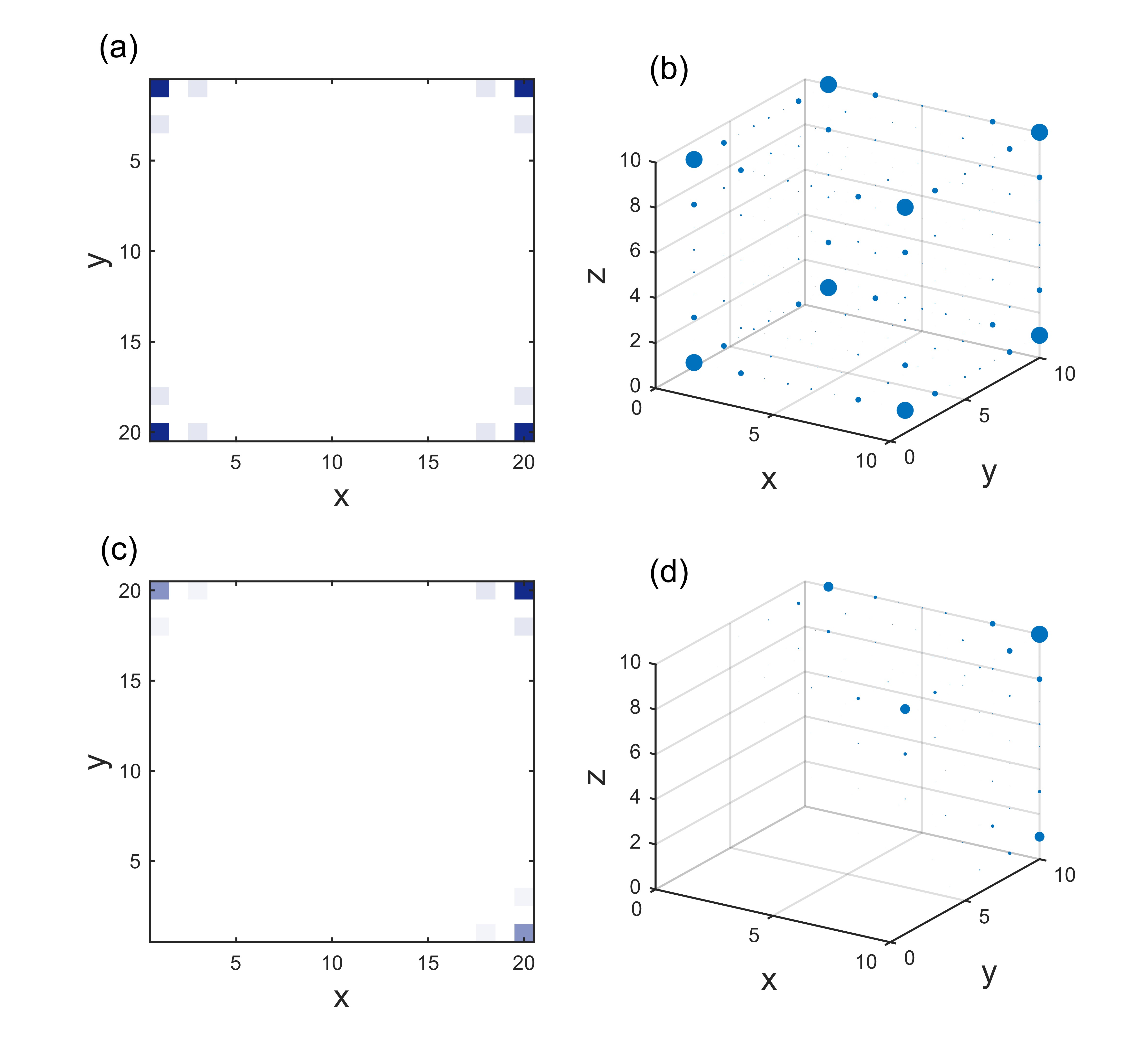}
  \caption{\label{fig_Appendix:FIG_Appendix_corner_distribution_modified}
  Distribution of the zero-energy corner state in (a,c) 2D and (b,d) 3D BBH model.
  In (a-b), the system is under OBC, corresponding to the CTBC with $r=0$.
  The corner states appear at the each corner, and the number of these corner states is respectively $N_{corner}^{2D} = 4$ and $N_{corner}^{3D} = 8$.
  In (c-d), the system is under a modified CTBC with $r=0$ instead of the OBC, and only one of the corner hosts the corner state.
  In both (c) and (d), the number of corner states is the same: ${N'}_{corner}^{2D} = {N'}_{corner}^{3D} = 2$.
  Other parameters are chosen as $\gamma = 0.5, \lambda = 1.5$.
    }
\end{figure}

\subsubsection{Modified CTBC and Winding number}
\begin{figure}[htp]
\centering
  \includegraphics[width = 0.8\columnwidth ]{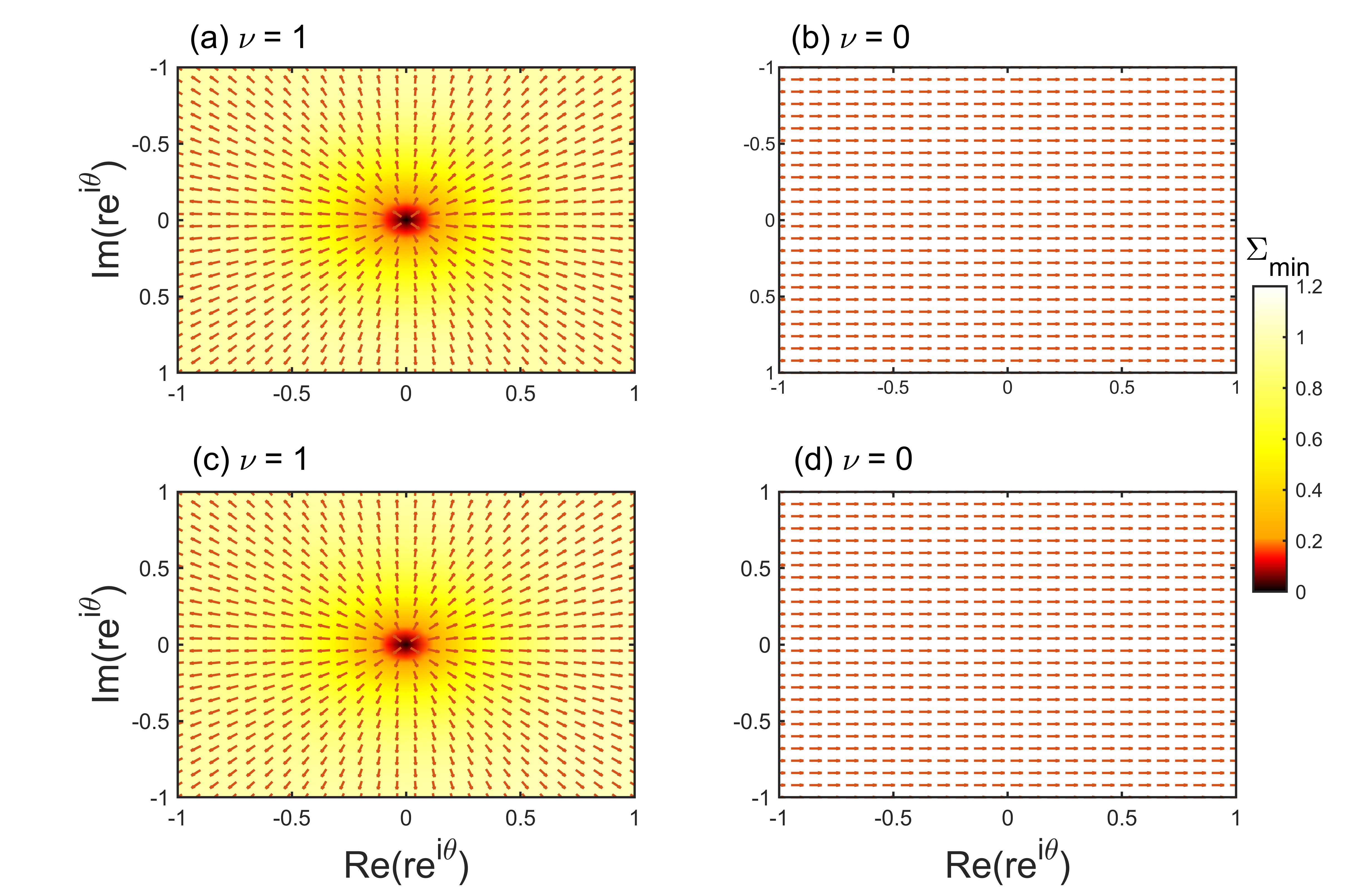}
  \caption{\label{fig_Appendix:FIG_Appendix_theta_r_diagram}
  Diagram of lowest singular values $\min(\Sigma)$ of the matrix $h(r, \theta)$ in: (a-b) 2D BBH model and (c-d) 3D BBH model within the modified CTBC situation.
  Arrows represent the angle of $\det h(r, \theta)$.
  The parameters are chosen to: $\gamma = 0.5$, $\lambda = 1.5$ in (a,c), while $\gamma = 1.5$, $\lambda = 0.5$ in (b,d).
    }
\end{figure}

Here, we consider the modified CTBC shown in Fig.~\ref{fig_Appendix:FIG_Appendix_Half_CTBC_illustration}, which is slightly different from the main text.
We shall show that the topological property of chiral-symmetric topological insulator in 2D and 3D can be also related to the winding number defined through the modified CTBC.
In this modified CTBC, all corners are still connected as the CTBC introduced in the main text.
This definition is closer to the CTBC originally proposed in Ref.~\citep{PhysRevLett.128.246602}.
However, only those tunnelings related to one of the corner is addressed with gauge field (twist angle) and the $r$ parameter.
Thus, $r=0$ does not correspond to the fully open boundary condition, but correspond to a partially open boundary condition, with only one of the corners left to be opened.
In a non-trivial phase, the corner state only appear at one of the corner, and the number of zero-energy states in 2D and 3D case is 2.
To make it easier to follow, we present the density distribution of the corner state under OBC and the modified CTBC with $r=0$ in Fig.~\ref{fig_Appendix:FIG_Appendix_corner_distribution_modified}.
With the modified CTBC, the winding number can be defined in a manner similar to the main text:
\begin{eqnarray}
\label{eqn:appendix_winding_number_z_complex}
\nu'  &=& \oint\limits_{\left( {r,\theta } \right) \in C} {z'{{\left( {r,\theta } \right)}^{ - 1}} \mathrm{d} z'\left( {r,\theta } \right)} ,\quad z' \equiv \det h'(r,\theta) \nonumber \\
&=& \frac{1}{{2\pi i}} \oint\limits_{\left( {r,\theta } \right) \in C} {{\rm{d}}\theta \;{\rm{Tr}}\left[ {h'{{\left( {r,\theta } \right)}^{ - 1}}{\partial _\theta }h'\left( {r,\theta } \right)} \right]} .
\end{eqnarray}
We apply this method to numerically compute $\det h'(r,\theta)$, as shown in Fig.~\ref{fig_Appendix:FIG_Appendix_theta_r_diagram}.
It can be seen that, for $r\ne 0$, the winding number for the non-trivial phase in 2D and 3D are both $\nu' = 1$, implying that only one pair of zero-energy modes occurs at $r=0$.
This clearly shows that the winding number defined in Eq.~\eqref{eqn:appendix_winding_number_z_complex} and the number of zero-energy modes under the modified CTBC are consistent.
This is confirmed numerically.

\subsubsection{Multipole chiral number and real-space formula of winding number}
\begin{figure}[htp]
\centering
  \includegraphics[width = 0.8\columnwidth ]{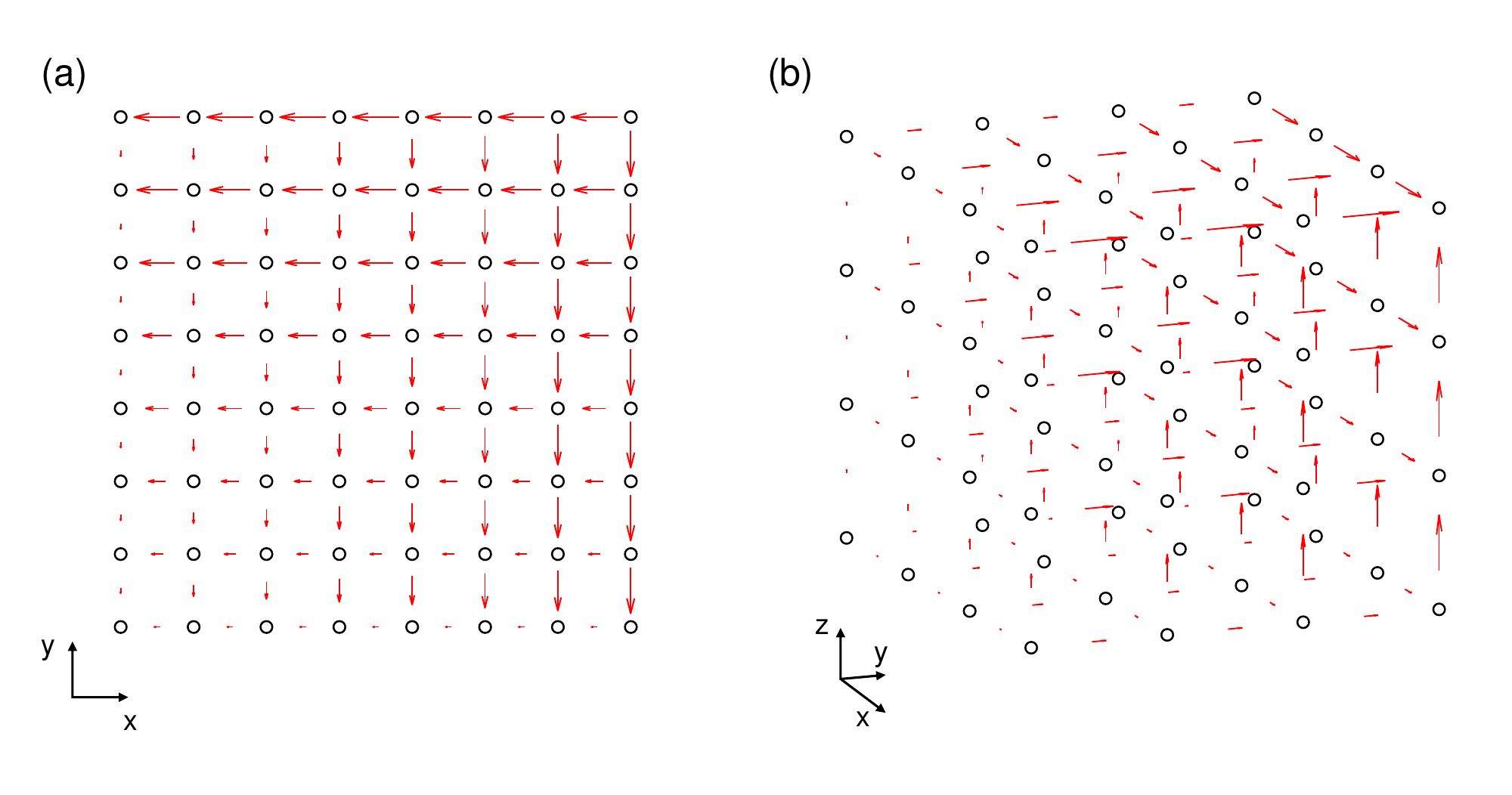}
  \caption{\label{fig_Appendix:FIG_Appendix_gauge_field_XYZ}
  Distribution of the gauge field after transformation from the reduced corner TBC using the (a) quadrupole operator and (b) octupole operator.
  The gauge field crossing the corner is omitted for simplicity.
   }
\end{figure}

Next, we derive the real-space formula of the winding number in Eq.~\eqref{eqn:appendix_winding_number_z_complex} and show that it is exactly the multipole chiral number.
With the multipole twist operators introduced in Eq.~\eqref{eqn:Appendix_twist_operators_Uxyz}, we perform the gauge transformation: $\tilde{H}'(\theta) = \hat{U}_{2\pi}\hat{H}(\theta)\hat{U}_{2\pi}^{-1}$ as we did in the main text.
The gauge field (twist angle) threaded to the system is then transformed to the bulk, as demonstrated in Fig.~\ref{fig_Appendix:FIG_Appendix_gauge_field_XYZ}.
The largest strength of the gauge field here is in the order of $1/L$ near the target corner, and therefore the twist angle in this scenario can be also treated as a perturbation.
Following the same method introduced in the main text, we can obtain the real-space formulas via the perturbative expansion for the 2D and 3D systems with some algebraic calculations:
\begin{eqnarray}
\label{eqn_appendix:nu_bott_index}
{\nu _{xy}} &=& \frac{1}{{2\pi i}}{\rm{Tr}}\left\{ {\log \left[ {h{'^{ - 1}}{\cal U}_{2\pi }^{xy}h'{{\left( {{\cal U}_{2\pi }^{xy}} \right)}^{ - 1}}} \right]} \right\}. \nonumber \\
{\nu _{xyz}} &=& \frac{1}{{2\pi i}}{\rm{Tr}}\left\{ {\log \left[ {h{'^{ - 1}}{\cal U}_{2\pi }^{xyz}h'{{\left( {{\cal U}_{2\pi }^{xyz}} \right)}^{ - 1}}} \right]} \right\},
\end{eqnarray}
with ${\cal U}_{2\pi }^{xy} = {\boldsymbol{\psi }}_A^\dag \hat U_{2\pi }^Q{{\boldsymbol{\psi }}_A},\;{\cal U}_{2\pi }^{xyz} = {\boldsymbol{\psi }}_A^\dag \hat U_{2\pi }^O{{\boldsymbol{\psi }}_A}$.
Note that the winding number can be expressed as the polarization difference between the two sublattices as mentioned in Eq.~\eqref{eqn:appendix_nu_nua_nub}.
We can also do it for the real-space formulas.
By using the flattened Hamiltonian and $q = {u_A}u_B^{ - 1}$, Eq.~\eqref{eqn_appendix:nu_bott_index} can be further expressed as
\begin{eqnarray}
\label{eqn_appendix:MCN_Q}
{\nu _{xy}} &=& \frac{1}{{2\pi i}}{\rm{Tr}}\left\{ {\log \left[ {q{'^{ - 1}}{\cal U}_{2\pi }^{xy}q'{{\left( {{\cal U}_{2\pi }^{xy}} \right)}^{ - 1}}} \right]} \right\} \nonumber \\
 &=& \frac{1}{{2\pi i}}{\rm{Tr}}\left\{ {\log \left[ {u{'_B}u{'_A}^{ - 1}{\cal U}_{2\pi }^{xy}u{'_A}u{'_B}^{ - 1}{{\left( {{\cal U}_{2\pi }^{xy}} \right)}^{ - 1}}} \right]} \right\} \nonumber \\
 &=& \frac{1}{{2\pi i}}{\rm{Tr}}\left\{ {\log \left[ {\left( {u{'_A}^{ - 1}{\cal U}_{2\pi }^{xy}u{'_A}} \right)\left( {u{'_B}^{ - 1}{{\left( {{\cal U}_{2\pi }^{xy}} \right)}^{ - 1}}u{'_B}} \right)} \right]} \right\} \nonumber \\
 &=& \frac{1}{{2\pi i}}{\rm{Tr}}\left\{ {\log \left[ {\bar Q_{xy}^A{{\left( {\bar Q_{xy}^B} \right)}^\dag }} \right]} \right\},
\end{eqnarray}
where we define $\bar Q_{xy}^{A,B} = {u'}_{A,B}^{ - 1}{\cal U}_{2\pi }^{xy}u{'_{A,B}}$.
As for the 3D case, we have a similar result:
\begin{equation}
\label{eqn_appendix:MCN_O}
{\nu _{xyz}} = \frac{1}{{2\pi i}}{\rm{Tr}}\left\{ {\log \left[ {\bar O_{xyz}^A{{\left( {\bar O_{xyz}^B} \right)}^\dag }} \right]} \right\},
\end{equation}
with $\bar O_{xyz}^{A,B} = {u'}_{A,B}^{ - 1}{\cal U}_{2\pi }^{xyz}u{'_{A,B}}$.
Eqs.~\eqref{eqn_appendix:MCN_Q} and \eqref{eqn_appendix:MCN_O} are exactly the real-space form of the winding number for chiral-symmetric HOTI proposed in Ref.~\citep{PhysRevLett.128.127601}.
Although the modified CTBC is quite different from the CTBC discussed in the main text, it is still capable to extract the existence of the zero-energy corner modes localized at one of the corners.
Provided that the system is symmetric under the rotation, i.e. the system is isotropic, then all corners are equivalent.
Thus, the winding number defined through the modified CTBC should be equivalent to the MWN defined through the CTBC in the main text, since they both characterize the zero-energy mode at the corner.

The real-space representation discussed above is the so-called Bott index.
Similarly, we can also derive the non-commutative form:
\begin{eqnarray}
\label{eqn_appendix:nu_non_commutative_q}
\nu'_{xy}  &=& \frac{1}{L^2}\mathrm{Tr}\left( {{h'}^{ - 1}}\left[ {{{Q_{xyz}}},{h'}} \right] \right), \nonumber \\
\nu'_{xyz}  &=& \frac{1}{L^3}\mathrm{Tr}\left( {{h'}^{ - 1}}\left[ {{{O_{xyz}}},{h'}} \right] \right),
\end{eqnarray}
which also provides a real-space method for bulk measurement in experiment, as already introduced in the main text.

\subsection{Higher-order finite difference for the numerical computation and the convergence of the BCH formula}
\subsubsection{Higher-order finite difference}
\begin{figure}
\centering
  \includegraphics[width = 1\columnwidth]{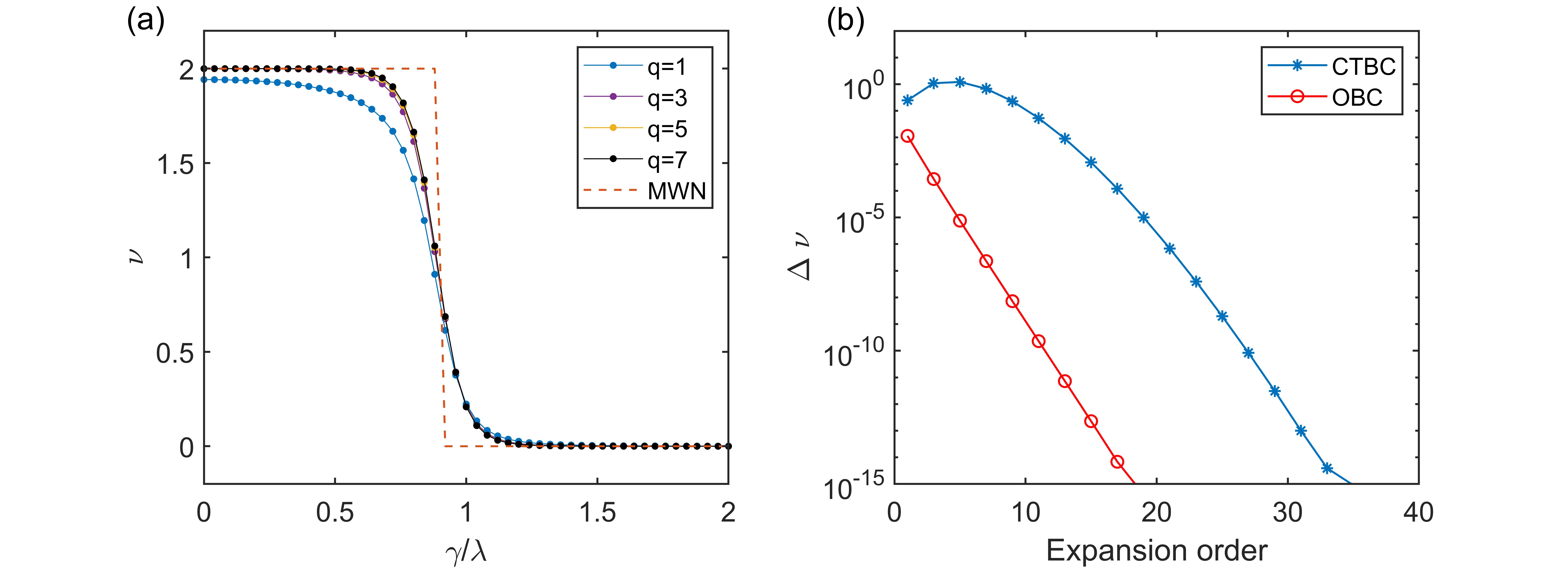}
  \caption{\label{fig_Appendix:FIG_Appendix_BCH_convergence}
   (a) Comparison between the MWN approximated from the $q$-order finite difference according to Eq.~\eqref{eqn_appendix:winding_number_BCH_expansion}.
   The dashed line is computed from the MWN defined through the CTBC introduced in the main text.
    (b) Difference between the MWN expanded by BCH formula and the original MWN as a function of expansion order according to Eq.~\eqref{eqn_appendix:BCH_difference} with $\gamma/\lambda = 1/3$.
  The size is chosen as $L=30$ in both plots.
   }
\end{figure}

In the main text, we have used the simplest finite difference to approximate the partial derivative:
\begin{equation}
{\left[ {{\partial _\theta }\tilde h\left( \theta  \right)} \right]_{\theta  = 0}} \approx \frac{{\tilde h\left( {\delta \theta } \right) - \tilde h\left( 0 \right)}}{{\delta \theta }},
\end{equation}
and then we let $\theta = 2\pi$ to attain the following useful expression
\begin{equation}
{\left[ {{\partial _\theta }\tilde h\left( \theta  \right)} \right]_{\theta  = 0}} \approx \frac{{{{\cal U}_{2\pi }}h{\cal U}_{2\pi }^{ - 1} - h}}{{2\pi }}.
\end{equation}
This approximation is valid only for sufficiently large systems, since the twist angle brings effects to the system in $1/L$ order.
For finite systems, the numerical result is quite sensitive to the accuracy of the partial derivative, and therefore this simple approximation is inadequate to produce a desirable (nearly) quantized MWN.
Particularly, near the phase transition point, where the spectral gap is extremely small, the perturbative expansion to the first order of $1/L$ is not enough.
The promote the accuracy, one can employee a higher-order approximation for the partial derivative.
Consider the expansion around $\theta = 0$:
\begin{equation}
\tilde h\left( \theta  \right) \approx h\left( 0 \right) + \sum\limits_{n = 1}^\infty  {{{\left[ {\frac{{{\partial ^n}}}{{\partial {\theta ^n}}}\tilde h\left( \theta  \right)} \right]}_{\theta  = 0}}\frac{{{\theta ^n}}}{{n!}}} ,
\end{equation}
we can approximate the first-order term by invoking the following linear superposition up to $q\in \mathbb{Z}$ terms:
\begin{equation}
{\left[ {{\partial _\theta }\tilde h\left( \theta  \right)} \right]_{\theta  = 0}} = \sum\limits_{m = 1}^q {\frac{\chi _m}{2}\left[ {\tilde h\left( {m\theta } \right) - \tilde h\left( { - m\theta } \right)} \right]}  + O\left( {\frac{1}{{{L^q}}}} \right).
\end{equation}
The coefficient $\chi_m$ can be solved by:
\begin{equation}
\left( {\begin{array}{*{20}{c}}
{{\chi _1}}\\
{{\chi _2}}\\
 \vdots \\
{{\chi _q}}
\end{array}} \right) = {\left( {\begin{array}{*{20}{c}}
1&2& \cdots &q\\
{{1^3}}&{{2^3}}& \cdots &{{q^3}}\\
 \vdots & \vdots & \ddots & \vdots \\
{{1^{2q - 1}}}&{{2^{2q - 1}}}& \cdots &{{q^{2q - 1}}}
\end{array}} \right)^{ - 1}}\left( {\begin{array}{*{20}{c}}
1\\
0\\
 \vdots \\
0
\end{array}} \right).
\end{equation}
Again, we let $\theta = 2\pi$ and use the relation
\begin{eqnarray}
\tilde h\left( {2m\pi } \right) &=& {{\cal U}_{2m\pi }}h{\cal U}_{2m\pi }^{ - 1} \nonumber \\
\tilde h\left( { - 2m\pi } \right) &=& {{\cal U}_{ - 2m\pi }}h{\cal U}_{ - 2m\pi }^{ - 1}\nonumber \\
 &=& {\cal U}_{2m\pi }^{ - 1}h{{\cal U}_{2m\pi }}
\end{eqnarray}
finally we obtain an efficient method to approximate the partial derivative:
\begin{equation}
{\left[ {{\partial _\theta }\tilde h\left( \theta  \right)} \right]_{\theta  = 0}} = \sum\limits_{m = 1}^q {{\chi _m}\left[ {{{\cal U}_{2m\pi }}h{\cal U}_{2m\pi }^{ - 1} - {\cal U}_{2m\pi }^{ - 1}h{{\cal U}_{2m\pi }}} \right]}  + O\left( {\frac{1}{{{L^q}}}} \right).
\end{equation}
This is particularly useful for numerical computation in finite systems.
Then, we show a numerical result by using different order of the finite difference in Fig.~\ref{fig_Appendix:FIG_Appendix_BCH_convergence} (a).
As we increase the order $q$ of the finite difference, the results tends to be well quantized away from the phase transition point.
The result near the phase transition point also becomes sharper for higher orders.
Besides,it should be noted that the perturbative expansion is poor near the phase transition point because the spectral gap is extremely small in this regime.
Thus, the real-space formula become non-quantized near the phase transition point as well.

\subsubsection{Convergence of BCH formula}
In the main text, the non-commutative form is derived by using the BCH formula up to the lowest order, which only works well under the OBC.
Here, we give a detailed description.
The BCH formula manifests that
\begin{equation}
{e^{i\frac{{2\pi }}{L}P}}h{e^{ - i\frac{{2\pi }}{L}P}} = h + \sum\limits_m {{{\left( {\frac{{2\pi i }}{L}} \right)}^m}\frac{1}{{k!}}{{\left[ {P,h} \right]}^m}} ,
\end{equation}
where 
\begin{equation}
{\left[ {P,h} \right]^m} \equiv \underbrace {\left[\cdots {P,\left[ {P,\left[ {P,h} \right]} \right]} \right] }_m  
\end{equation}
denotes the iterative commutation for $m$ times.
The MWN can be therefore approximated via
\begin{equation}
\label{eqn_appendix:winding_number_BCH_expansion}
\nu  \approx \frac{1}{2\pi} \mathrm{Im}{\mathrm{Tr}}\left\{ {{h^{ - 1}}\left[ {h + \sum\limits_m {{{\left( {\frac{{2\pi i}}{L}} \right)}^m}\frac{1}{{k!}}{{\left[ {P,h} \right]}^m}} } \right]} \right\}.
\end{equation}
Note that the even-order term is purely real, and therefore it contributes nothing to the MWN.
Now, we examine the performance of this expansion under CTBC and OBC for 2D system by computing the difference between the expansions and the MWN approximated by simple finite difference ${\rm{Tr}}\left( {{h^{ - 1}}{e^{i\frac{{2\pi }}{L}F}}h{e^{ - i\frac{{2\pi }}{L}P}} - I} \right) $:
\begin{equation}
\label{eqn_appendix:BCH_difference}
\Delta \nu  =\frac{1}{2\pi} {\mathop{\rm Im}\nolimits} \left| {{\rm{Tr}}\left\{ {{h^{ - 1}}\left[ {h + \sum\limits_m {{{\left( {\frac{{2\pi }}{L}} \right)}^m}\frac{1}{{m!}}{{\left[ {P,h} \right]}^m}} } \right]} \right\} - {\rm{Tr}}\left( {{h^{ - 1}}{e^{i\frac{{2\pi }}{L}F}}h{e^{ - i\frac{{2\pi }}{L}P}} - I} \right)} \right|,
\end{equation}
as shown in Fig.~\ref{fig_Appendix:FIG_Appendix_BCH_convergence} (b).
For OBC, we have explicitly ruled out zero singular values and use the pseudo-inverse formalism for the matrix $h$.
It can be seen that the BCH formula is exact if the expansion is made for sufficient orders.
However, the convergence is slow under the CTBC.
This can be understood by noticing that eigenstates are generally continuous between corners due to the CTBC. 
The poor performance of the BCH formula can be attributed to the discontinuity of the operators $P$ near the corner. 
Under the OBC, this issue is reduced and therefore the convergence become promising.
Therefore, it is adequate to approximate the MWN by only keeping the first order in experiments under the OBC.

\begin{figure}
\centering
  \includegraphics[width = 1\columnwidth ]{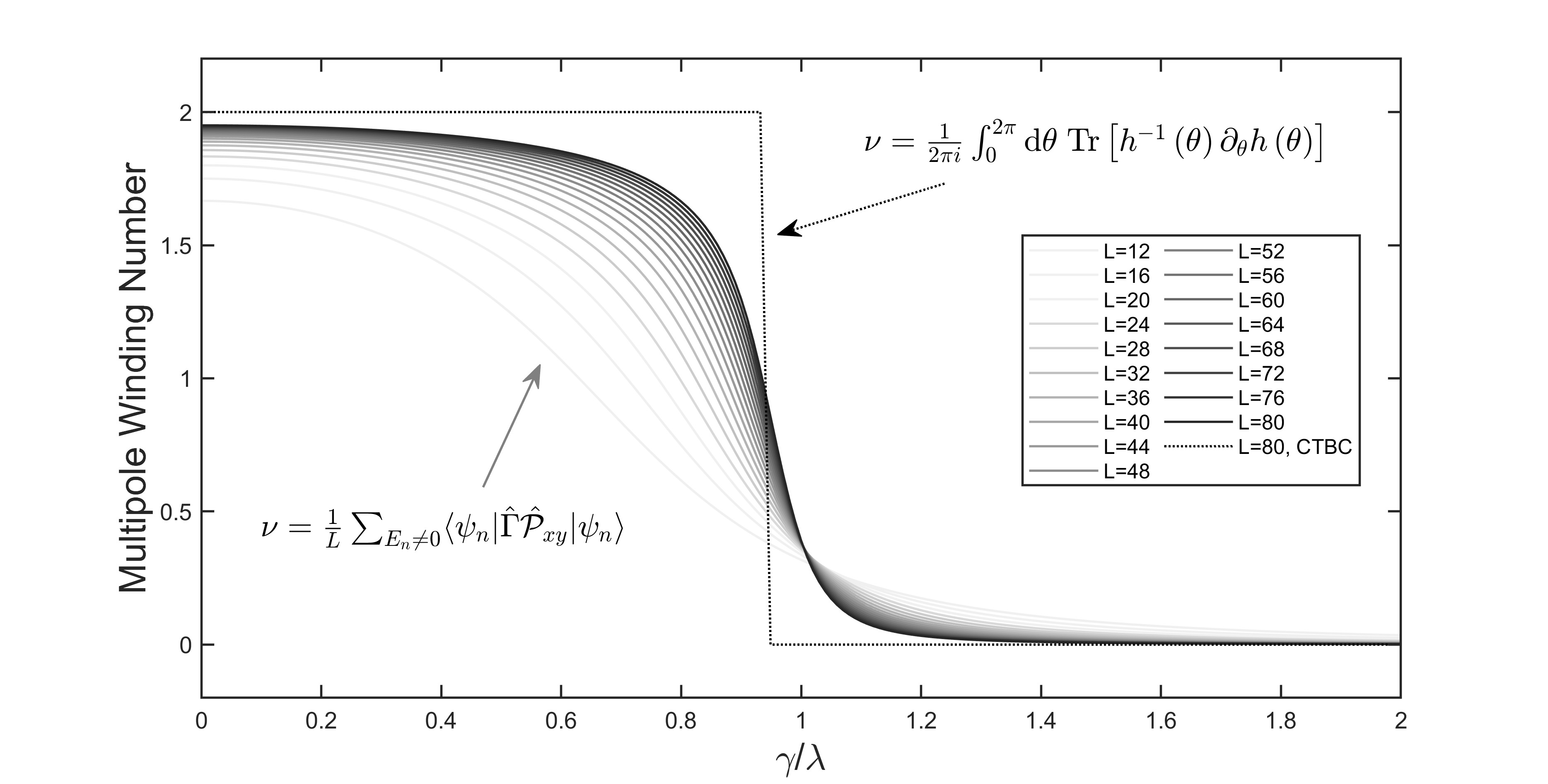}
  \caption{\label{fig_Appendix:FIG_Appendix_WindingNumber_OBC_scaling}
  Non-commutative form of the MWN [Eq.~\eqref{eqn_appendix:nu_Gamma_P_expression_derivation}] as a function of $\gamma/\lambda$ with various system's size (solid lines).
  For comparison, we also plot the MWN defined through the CTBC for $L=80$ system (dashed line).
  These results are computed based on the 2D BBH model.
   }
\end{figure}

\subsection{Derivation of the formula for real-space measurement from the non-commutative form}
The non-commutative form of the MWN is derived in a matrix notation in the main text: $\nu=\frac{1}{L}\mathrm{Tr}(q^{-1}[P,q])$.
Here, we show how to derive the formula for experimental measurement $\nu  = \frac{1}{L}\sum\limits_{{E_n} \ne 0} {\langle {\psi _n}|\hat \Gamma \hat {\cal P}|{\psi _n}\rangle } $ based on this non-commutative form.
Substituting $q\equiv {\boldsymbol{\psi }}_A^\dag \hat Q{{\boldsymbol{\psi }}_B}$, $q^{-1}\equiv {\boldsymbol{\psi }}_B^\dag \hat Q{{\boldsymbol{\psi }}_A}$ and $P \equiv {\boldsymbol{\psi }}_A^\dag \hat {\cal P}{{\boldsymbol{\psi }}_A} = {\boldsymbol{\psi }}_B^\dag \hat {\cal P}{{\boldsymbol{\psi }}_B}$, there is
\begin{eqnarray}
\label{eqn_appendix:real_space_winding_num_Qab_Qba}
\nu  &=& \frac{1}{L}{\rm{Tr}}\left( {{q^{ - 1}}\left[ {P,q} \right]} \right) \nonumber \\
 &=& \frac{1}{L}{\rm{Tr}}\left( {{\boldsymbol{\psi }}_B^\dag \hat Q{{\boldsymbol{\psi }}_A}\left[ {{\boldsymbol{\psi }}_A^\dag \hat {\cal P}{{\boldsymbol{\psi }}_A},{\boldsymbol{\psi }}_A^\dag \hat Q{{\boldsymbol{\psi }}_B}} \right]} \right) \nonumber \\
 &=& \frac{1}{L}{\rm{Tr}}\left( {{\boldsymbol{\psi }}_B^\dag \hat Q{{\boldsymbol{\psi }}_A}{\boldsymbol{\psi }}_A^\dag \hat {\cal P}{{\boldsymbol{\psi }}_A}{\boldsymbol{\psi }}_A^\dag \hat Q{{\boldsymbol{\psi }}_B} - {\boldsymbol{\psi }}_B^\dag \hat Q{{\boldsymbol{\psi }}_A}{\boldsymbol{\psi }}_A^\dag \hat Q{{\boldsymbol{\psi }}_B}{\boldsymbol{\psi }}_B^\dag \hat {\cal F}{{\boldsymbol{\psi }}_B}} \right) \nonumber \\
 &=& \frac{1}{L}{\rm{Tr}}\left( {{{\hat \Gamma}_B}\hat Q{{\hat \Gamma}_A}\hat {\cal F}{{\hat \Gamma}_A}\hat Q{{\hat \Gamma}_B} - {{\hat \Gamma}_B}\hat Q{{\hat \Gamma}_A}\hat Q{{\hat \Gamma}_B}\hat {\cal P}} \right) \nonumber \\
 &=& \frac{1}{L}{\rm{Tr}}\left( {{{\hat Q}_{BA}}\hat {\cal P}{\hat{Q}_{AB}} - {{\hat Q}_{BA}}{{\hat Q}_{AB}}\hat {\cal P}} \right) \nonumber \\
 &=& \frac{1}{L}{\rm{Tr}}\left( {{{\hat Q}_{BA}}\left[ {\hat {\cal P},{\hat{Q}_{AB}}} \right]} \right),
\end{eqnarray}
in which we denote ${\hat Q_{AB}} = {{\hat \Gamma}_A}\hat Q{{\hat \Gamma}_B}$ and ${\hat Q_{BA}} = {{\hat \Gamma}_B}\hat Q{{\hat \Gamma}_A}$, with $\hat{\Gamma}_{A,B} = \boldsymbol{\psi}_{A,B}\boldsymbol{\psi}_{A,B}^\dag$ being the projector onto $A,B$ sector.
The flattened Hamiltonian $\hat{Q}$ can be expressed as $\hat{Q} = \hat{Q}_{AB} + \hat{Q}_{BA}$.
Then, it can be found that
\begin{eqnarray}
{\hat Q_{AB}} &=& {{\hat \Gamma}_A}\hat Q{{\hat \Gamma}_B} \nonumber \\
 &=& {{\hat \Gamma}_A}\left( {\sum\limits_{{E_n} > 0} {|{\psi _n}\rangle \langle {\psi _n}|}  - \sum\limits_{{E_n} < 0} {|{\psi _n}\rangle \langle {\psi _n}|} } \right){{\hat \Gamma}_B} \nonumber  \\
 &=& {{\hat \Gamma}_A}\left( {\sum\limits_{{E_n} < 0} {\left( {\hat \Gamma |{\psi _n}\rangle \langle {\psi _n}|\hat \Gamma } \right)}  - \sum\limits_{{E_n} < 0} {|{\psi _n}\rangle \langle {\psi _n}|} } \right){{\hat \Gamma}_B} \nonumber  \\
 &=& 2{{\hat \Gamma}_A}\sum\limits_{{E_n} < 0} {\left( {|{\psi _n}\rangle \langle {\psi _n}|} \right)} {{\hat \Gamma}_B},
\end{eqnarray}
where we have used $\hat \Gamma  = {{\hat \Gamma}_A} - {{\hat \Gamma}_B}$ and ${\left( {{{\hat \Gamma}_{A,B}}} \right)^2} = {{\hat \Gamma}_{A,B}}$.
Similarly, there is 
\begin{equation}
{\hat Q_{BA}} = 2{{\hat \Gamma}_B}\sum\limits_{{E_n} < 0} {\left( {|{\psi _n}\rangle \langle {\psi _n}|} \right)} {{\hat \Gamma}_A}.
\end{equation}
Hence, Eq.~\eqref{eqn_appendix:real_space_winding_num_Qab_Qba} can be expanded to
\begin{eqnarray}
\label{eqn_appendix:nu_Gamma_P_expression_derivation}
\nu  &=& \frac{1}{L}{\rm{Tr}}\left( {{{\hat Q}_{BA}}\left[ {\hat{\cal P},{{\hat Q}_{AB}}} \right]} \right) \nonumber \\
 &=&\frac{4}{L}{\rm{Tr}}\left( {{{\hat \Gamma}_B}\sum\limits_{{E_n} < 0} {\left( {|{\psi _n}\rangle \langle {\psi _n}|} \right)} {{\hat \Gamma}_A}\left[ {{{\hat \Gamma}_A}\sum\limits_{{E_n} < 0} {\left( {|{\psi _n}\rangle \langle {\psi _n}|} \right)} {{\hat \Gamma}_B}, \hat{\cal P}} \right]} \right) \nonumber\\
 &=&\frac{4}{L}{\rm{Tr}}\left( {{{\hat \Gamma}_B}\sum\limits_{{E_n} < 0} {\left( {|{\psi _n}\rangle \langle {\psi _n}|} \right)} {{\hat \Gamma}_A} \hat{\cal P} {{\hat \Gamma}_A}\sum\limits_{{E_n} < 0} {\left( {|{\psi _n}\rangle \langle {\psi _n}|} \right)} {{\hat \Gamma}_B} - {{\hat \Gamma}_B}\sum\limits_{{E_n} < 0} {\left( {|{\psi _n}\rangle \langle {\psi _n}|} \right)} {{\hat \Gamma}_A}\sum\limits_{{E_n} < 0} {\left( {|{\psi _n}\rangle \langle {\psi _n}|} \right)} {{\hat \Gamma}_B}\hat{\cal P}} \right) \nonumber\\
 &=&\frac{4}{L}{\rm{Tr}}\left( {{{\hat \Gamma}_B}\sum\limits_{{E_n} < 0} {\left( {|{\psi _n}\rangle \langle {\psi _n}|} \right)} {{\hat \Gamma}_A}\hat{\cal P}{{\hat \Gamma}_A}\sum\limits_{{E_n} < 0} {\left( {|{\psi _n}\rangle \langle {\psi _n}|} \right)} {{\hat \Gamma}_B} - {{\hat \Gamma}_A}\sum\limits_{{E_n} < 0} {\left( {|{\psi _n}\rangle \langle {\psi _n}|} \right)} {{\hat \Gamma}_B}\hat{\cal P}{{\hat \Gamma}_B}\sum\limits_{{E_n} < 0} {\left( {|{\psi _n}\rangle \langle {\psi _n}|} \right)} {{\hat \Gamma}_A}} \right) \nonumber\\
 &=& \frac{2}{L}\sum\limits_{{E_n} < 0} {\langle {\psi _n}|\left( {{{\hat \Gamma}_A}\hat{\cal P}{{\hat \Gamma}_A} - {{\hat \Gamma}_B}\hat{\cal P}{{\hat \Gamma}_B}} \right)|{\psi _n}\rangle } \nonumber \\
 &=&\frac{1}{L}\sum\limits_{{E_n} \ne 0} {\langle {\psi _n}|\hat \Gamma \hat{\cal P}|{\psi _n}\rangle } ,
\end{eqnarray}
where we have used the following relations in chiral-symmetric systems:
\begin{eqnarray}
\left[ {{{\hat \Gamma}_{A,B}},\hat{\cal P}} \right] &=& 0 \nonumber \\
\langle {\psi _n}|{{\hat \Gamma}_A}|{\psi _{n'}}\rangle  &=& \frac{1}{2}{\delta _{n, - n'}} \nonumber \\
\langle {\psi _n}|{{\hat \Gamma}_B}|{\psi _{n'}}\rangle  &=&  - \frac{1}{2}{\delta _{n, - n'}},
\end{eqnarray}
in which $-n$ implies the index of the eigenstate with opposite energy $E_{-n} = -E_n$.
In the last row of Eq.~\eqref{eqn_appendix:nu_Gamma_P_expression_derivation}, we have summed over all eigenstates except for the zero-energy terms, which is beneficial for the experimental measurement.
In the thermodynamic limit, this lowest-order approximation is expected to be accurate, and the non-commutative method will align with the MWN from the CTBC.
To verify it, we numerically compute the non-commutative form of the MWN in Eq.~\eqref{eqn_appendix:nu_Gamma_P_expression_derivation} for 2D BBH model as a function of $\gamma/\lambda$ for different sizes, as shown in Fig.~\ref{fig_Appendix:FIG_Appendix_WindingNumber_OBC_scaling}.
It can be seen that the result shows a tendency to the exact MWN in the thermodynamic limit.
%
% Also, as the results at two sides of the phase transition point $\gamma/\lambda=1$ will respectively flow to $\nu=2$ or $\nu=0$, the intersecting point among these lines may hint the real phase transition point ($\gamma/\lambda = 1$) under the PBC.
%%

\subsection{Discussions on the measurement via dynamical evolution}
\begin{figure}
\centering
  \includegraphics[width = 1\columnwidth ]{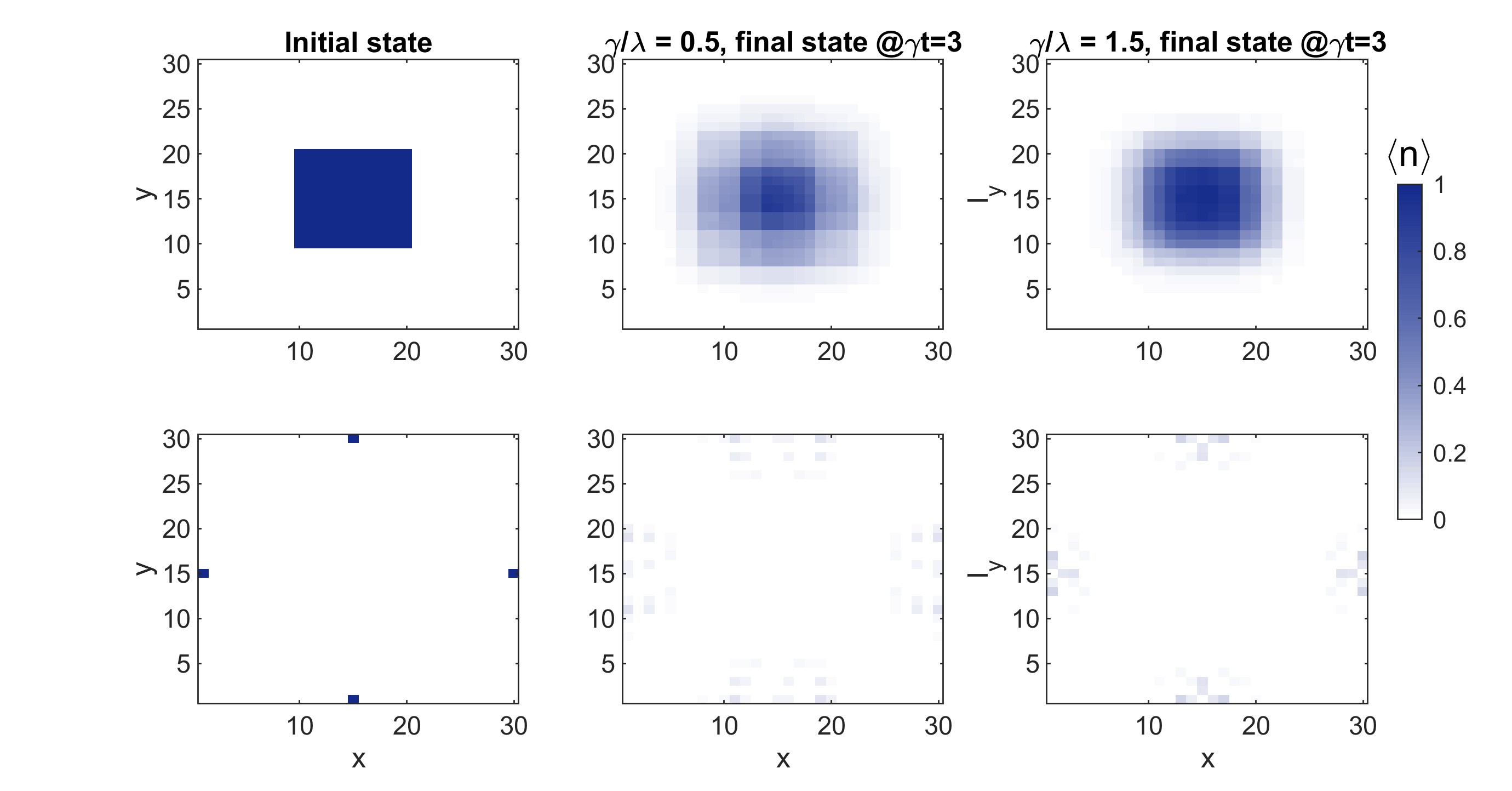}
  \caption{\label{fig_Appendix:FIG_initial_final_distribution}
 Ensemble density distribution of the initial state (first column) and final state (second and third columns).
 The first and second rows respectively correspond to the bulk and edge configuration.
 System's size is set to $L=30$.
   }
\end{figure}

As mentioned in the main text, the MWN can be expressed as the expectation value of the chiral multipole polarization [Eq.~\eqref{eqn_appendix:nu_Gamma_P_expression_derivation}].
To measure it experimentally, we can choose appropriate initial states that cover all eigenstates with non-zero energy and let it evolves:
\begin{equation}
|\Psi \left( t \right)\rangle  = \sum\limits_{{E_n} \ne 0} {{e^{ - i{E_n}t}}{c_n}|{\psi _n}\rangle } .
\end{equation}
The time-dependent expectation of some operators $\hat{O}$ can be written as 
\begin{equation}
\label{eqn_appendix:O_evolution_measurement_off_diagonal}
\langle \Psi \left( t \right)|\hat O|\Psi \left( t \right)\rangle  = \sum\limits_{{E_n} \ne 0} {{e^{ - i\left( {{E_n} - {E_{n'}}} \right)t}}{c_n}{c_{n'}}^*\langle {\psi _{n'}}|\hat O|{\psi _n}\rangle } .
\end{equation}
If we average it for sufficiently long time $\Delta t\to\infty$, the off-diagonal terms $(n'\ne n)$ will vanish:
\begin{equation}
\bar O = \frac{1}{{\Delta t}}\int_t^{t + \Delta t} {dt\;\langle \Psi \left( t \right)|\hat O|\Psi \left( t \right)\rangle }  \approx \sum\limits_{{E_n} \ne 0} {|{c_n}{|^2}\langle {\psi _{n}}|\hat O|{\psi _n}\rangle } .
\end{equation}
This provides an effective method to measure the MWN in experiment, as mentioned in the main text.
It can be seen that the most important step is to choose appropriate initial states.
In 2D BBH model with open boundary, it turns out that there are bulk states and edge states with non-zero energy when the system is non-trivial.
The contributions from bulk states and the edge states are different.
Therefore, we should take them both into account in the experiment.
In this work, we consider an ensemble of Fock states as the initial state, which can be conveniently realized in many platforms.
In experiment, one can use appropriate Fock states to evolve and average the measurement to obtain the desired result.
To ensure that these initial states can uniformly cover the bulk states and edge states, we let the Fock state distribute at the center of the bulk and the center of the edge, as illustrated in the first column of Fig.~\ref{fig_Appendix:FIG_initial_final_distribution}.
This generally makes the Fock state overlap uniformly with the bulk state and edge state, enabling the ensemble average of the MWN measurement.

Near the phase transition point, the localization length of the corner state becomes considerably large, and therefore the Fock state at the center of the bulk or the edge will inevitably have overlap with the zero-energy corner state.
Since the measurement of the MWN requires the exclusion of zero-energy corner states, this will cause inaccurate measurement.
To reduce this effect, we consider a relatively short evolution for these Fock states such that they will not reach the boundary, as shown in the second and third column of Fig.~\ref{fig_Appendix:FIG_initial_final_distribution}.
This can avoid the effect of corner state while give a reasonably correct measurement although the evolution time is relatively short and the off-diagonal term in Eq.~\eqref{eqn_appendix:O_evolution_measurement_off_diagonal} may not be well eliminated.
Physically, this is because the short-time evolution can still capture the topological information of the regime inside the bulk or edge where states go through before reaching the corner.
In experiment, the relatively short time evolution beneficial since the coherent time of the evolving quantum state may be very short.

% \bibliography{HOTI_real_space_BIB}
